\documentclass[journal]{IEEEtran}
\usepackage[latin1]{inputenc}
\usepackage{booktabs} 
\usepackage{amsmath}
\usepackage{amssymb}
\usepackage{graphicx}
\usepackage{caption}
\usepackage{setspace}
\usepackage{array}
\usepackage{tabu}
\usepackage{nomencl}
\usepackage{moreverb}
\usepackage{color}

\ifCLASSINFOpdf
\else
\fi

\begin{document}

\title{Emerging Privacy Issues and Solutions in Cyber-Enabled Sharing Services: From Multiple Perspectives}

\author{Ke~Yan, {\it Member}, {\it IEEE}, Wen~Shen, Qun~Jin, {\it Senior Member}, {\it IEEE}, Huijuan~Lu
\thanks{$^*$Corresponding author: Qun Jin (Email address: jin@waseda.jp).
Author's addresses: K. Yan, H. Lu {and} Q. Jin, College of Information Engineering, China Jiliang University, 258 Xueyuan Street, Hangzhou, China, 310018, emails: yanke@cjlu.edu.cn, keddiyan@gmail.com (K. Yan) and hjlu@cjlu.edu.cn (H. Lu); W. Shen, Department of Informatics, University of California Irvine, Irvine, CA 92697, email: wen.shen@uci.edu;  Q. Jin, Department of Human Informatics and Cognitive Sciences, Waseda University, 2-579-15 Mikajima, Tokorozawa, 359-1192, Japan.}}
\maketitle

\begin{abstract}
Fast development of sharing services has become a crucial part of the cyber-enabled world construction process, as sharing services reinvent how people exchange and obtain goods or services. However, privacy leakage or disclosure remains as a key concern during the sharing service development process. While significant efforts have been undertaken to address various privacy issues in recent years, there is a surprising lack of review for privacy concerns in the cyber-enabled sharing world. To bridge the gap, in this study, we survey and evaluate existing and emerging privacy issues relating to sharing services from various perspectives. Differing from existing similar works on surveying sharing practices in various fields, our work comprehensively covers six branches of sharing services in the cyber-enabled world, and selects solutions mostly from the recent five to six years. We conclude the issues and solutions from three perspectives, namely, from users', platforms' and service providers' perspectives. Hot topics and less discussed (cold) topics are identified, which provides hints to researchers for their future studies.
\end{abstract}


\IEEEpeerreviewmaketitle

\section{Introduction}

Cyberization is transforming our physical living world into a virtual computerized world by leveraging the Internet and computational methodologies~\cite{ma2016cybermatics, ma2016perspectives}. In the virtual computerized world, or more specifically, the cyber-enabled world, people are connected via Internet regardless of physical distances. Cyber-enabled sharing services, or in short, sharing services, which provide information, goods, and services in a shared form to multiple individuals, who know or do not know each other, are essential and necessary components of cyber-world development and probably the most exciting cyber-related concept in the current stage of cyberization. Sharing services encourage people to share both virtual and physical assets through the Internet using cyber-enabled clients, including mobile phones, all kinds of computers and similar digital devices. Sharing services contribute to the fast development of cyber technology, where the control, responsibility for the common good, earnings, capitalization, information,  and efforts are all shared among the participants or distributed to peer members~\cite{schor2014debating}. In recent years, cyberized sharing service companies, such as Uber, Airbnb, Etsy and Amazon Family Library, have been overwhelmingly popular and enjoyed incredible growth~\cite{zervas2014rise, belk2014you}.

There are various reasons for people to participate in sharing practices. For instance, no single entity or person can control the whole market or economy, although some participants have more regulatory power than others. All participants share the responsibility of making the market to operate healthily. This form of collaborative economy or peer-to-peer (P2P) sharing leads to more efficient resource allocations and more sustainable lifestyles. However, any participant in the sharing practice, regardless whether it is a user or service provider, can be a potential attacker who compromises legitimate users' privacy. Therefore, to attract more people to share, it is necessary to build trust, establish reputation, protect privacy and guarantee security for both the user and service provider~\cite{bella2011enforcing}. Personal privacy concern is the main factor that hinders the development of sharing services in the cyber-enabled world~\cite{du1985data,daswani2003open}. On one hand, people are reluctant to adopt sharing practices due to privacy disclosure concerns~\cite{feeney2015ridesharing,gobble2015regulating,li2015privacy}; on the other hand, sharing service providers insist that personal data is part of the necessary information in user experience analysis for improving service quality. While only privacy protection is explored in this paper, the authors would like to note that privacy is relevant and closely related to trust, reputation and security. Users need to trust the service provider, which implies that the service provider must have a good reputation that the users can trust. Reputations are established through the interactions between the users and service providers. However, during the interaction process, privacy issues arise, since private information pieces from both parties are inevitably revealed to each other \cite{Dimitriou2012Multi, Dimitriou2014Multi}.

Unfortunately, due to the fast development pace of sharing service technology, privacy issues were not well addressed before sharing services were widely spread over the physical world \cite{katz2015regulating}. For example, in the ridesharing service practice, although the business model exists for quite a while, there are still many privacy leakage concerns, including location privacy concerns, driver/customer's personal information leakage concerns, physical privacy concerns and etc. \cite{christin2016privacy}. Cyber-technologies that can be used to protect various aspects of privacy are urgently desired to prohibit both the user and service provider from revealing each other's sensitive information. In the starting stage of the sharing economy, some service providers intentionally neglect the privacy issues to survive in the highly competitive business environment. In other words, profit is usually the highest priority for most starter-level sharing service companies. In this study, we surveyed over one hundred research works from recent years that are closely related to the privacy issues with the newly developed sharing service technologies and observed that the privacy protection level is highly related to the number of users who participate in the sharing service, which affects the final profit of the service providers. In addition, from the user's perspective, increasing the self-awareness of privacy disclosure is an important task for the users to protect themselves in the current stage of cyberization.

In summary, the emerging privacy issues of sharing services in the cyber-enabled world and the available solutions are reviewed comprehensively. From the literature, we summarize the sharing services in the current stage of the cyber-enabled world into two categories \cite{schor2014debating, botsman2011s}:
\begin{itemize}
\item {\bf Crowdsourcing} employs collective intelligence or power to fulfil tasks or achieve goals. Concrete examples of crowdsourcing are Internet crowdsourcing marketplaces, crowdfunding,  and crowdtesting \cite{doan2011crowdsourcing}. For a typical crowdsourcing practice, there are, in general, three roles involved: the task requester, the platform and the worker. The task requester posts tasks on the platform and attracts workers to finish the job in a crowdsourcing way.
\item {\bf Collaborative consumption} allows consumers to use products or services without full ownership. Concrete examples of collaborative consumption include collaborative online shopping, ridesharing,  and homesharing practices~\cite{belk2014you}. For a typical collaborative consumption model, there are again three roles involved: the host, the platform and the customer. Differing from the crowdsourcing practice, the host provides P2P sharing of goods or services to customers through an online platform.
\end{itemize}

\begin{figure*}[h!t!b!]
  \centering
  \includegraphics[width=7.3in]{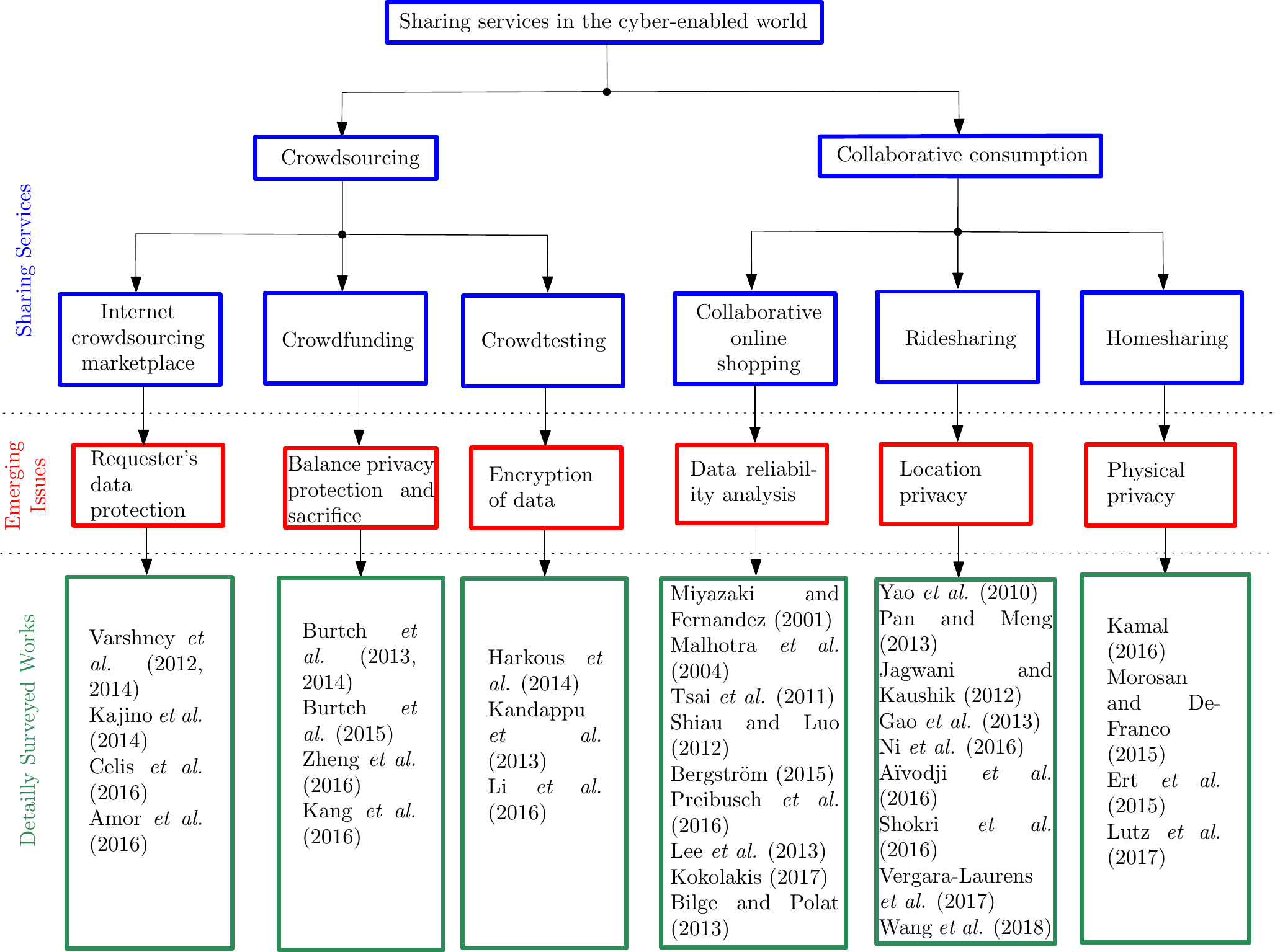}
    \caption{Taxonomy of sharing services in the cyber-enabled world following the categorization of crowdsourcing and collaborative consumption practices (in blue rectangles), with identified emerging privacy issues (in red rectangles) and surveyed works in the literature (in green rectangles).}
    \label{fig:overall2}
\end{figure*}

In this study, we refer to the combination of task requesters and hosts as service providers, and the combination of workers and customers as users. The review of privacy issues and solutions follows the above two outlines and reveals the main concerns in the literature, which include the requester's data protection, the balance between privacy protection and sacrifice, data encryption, unreliable data analysis, location privacy and physical privacy. Figure \ref{fig:overall2} lists a taxonomy of important works that are surveyed for privacy issues and solutions in crowdsourcing and collaborative consumption practices.

Although there are similar works concerning privacy in sharing practices from the literature, e.g., \cite{androutsellis2002survey, aggarwal2008general, fung2010privacy, rahman2015survey, heurix2015taxonomy}, they focused on traditional privacy protection methods. Traditional privacy protection techniques, including k-anonymity \cite{samarati2001protecting, sweeney2002k}, l-diversity \cite{machanavajjhala2007diversity} and t-closeness \cite{li2007t}, have been heavily reviewed in the past few decades. In contrast, our work focuses on privacy protection technique development in recent years, skips the traditional approaches and covers technologies comprehensively in the area of cyber-enabled sharing services. Most surveyed works in this study were published in past five to six years. The sources of the reviewed papers include the most popular databases, such as ACM Digital Library, IEEE Xplore Digital Library, Springer Link and ScienceDirect. The searched keywords include `sharing service', `privacy issue', `privacy protection`, `crowdsourcing privacy', `collaborative consumption privacy', `crowdfunding privacy' and etc.

The main contributions of this work include 1) categorizing recent research studies working on privacy issues of sharing services into trends, 2) identifying the hot/cold research topics, and 3) finding the research gaps for real-world sharing services to better protect people's privacy. For example, from Fig. \ref{fig:overall2}, we found that data reliability analysis and location privacy are two hot topics for collaborative consumption, whereas the physical privacy issue in homesharing practice is less discussed. Moreover, there are research works indicating that physical privacy is also largely concerned by users in the sharing service practices. Those less discussed topics require more attention in future studies.

The remaining parts of this work are organized as follows: The emerging privacy issues and solutions of crowdsourcing are analyzed in Section \ref{sec:Crowdsourcing}. The emerging privacy issues and solutions of collaborative consumption are reviewed in Section \ref{sec:collaborative}. In Section \ref{sec:emerging}, all six branches discussed in Sections \ref{sec:Crowdsourcing} and \ref{sec:collaborative} are summarized from three perspectives, namely, user, platform and service provider perspectives. Section \ref{sec:openIssues} raises open research issues for each branch of the sharing services and from the three perspectives mentioned in Section \ref{sec:emerging}. In Section \ref{sec:conclusion}, several conclusions are drawn regarding cyber technology development to predict the future trends in the development of cyber-enabled sharing technologies.

\section{Privacy Issues and Solutions in Crowdsourcing Practices}\label{sec:Crowdsourcing}

\begin{figure*}[h!t!b!]
  \centering
  \includegraphics[width=5.5in]{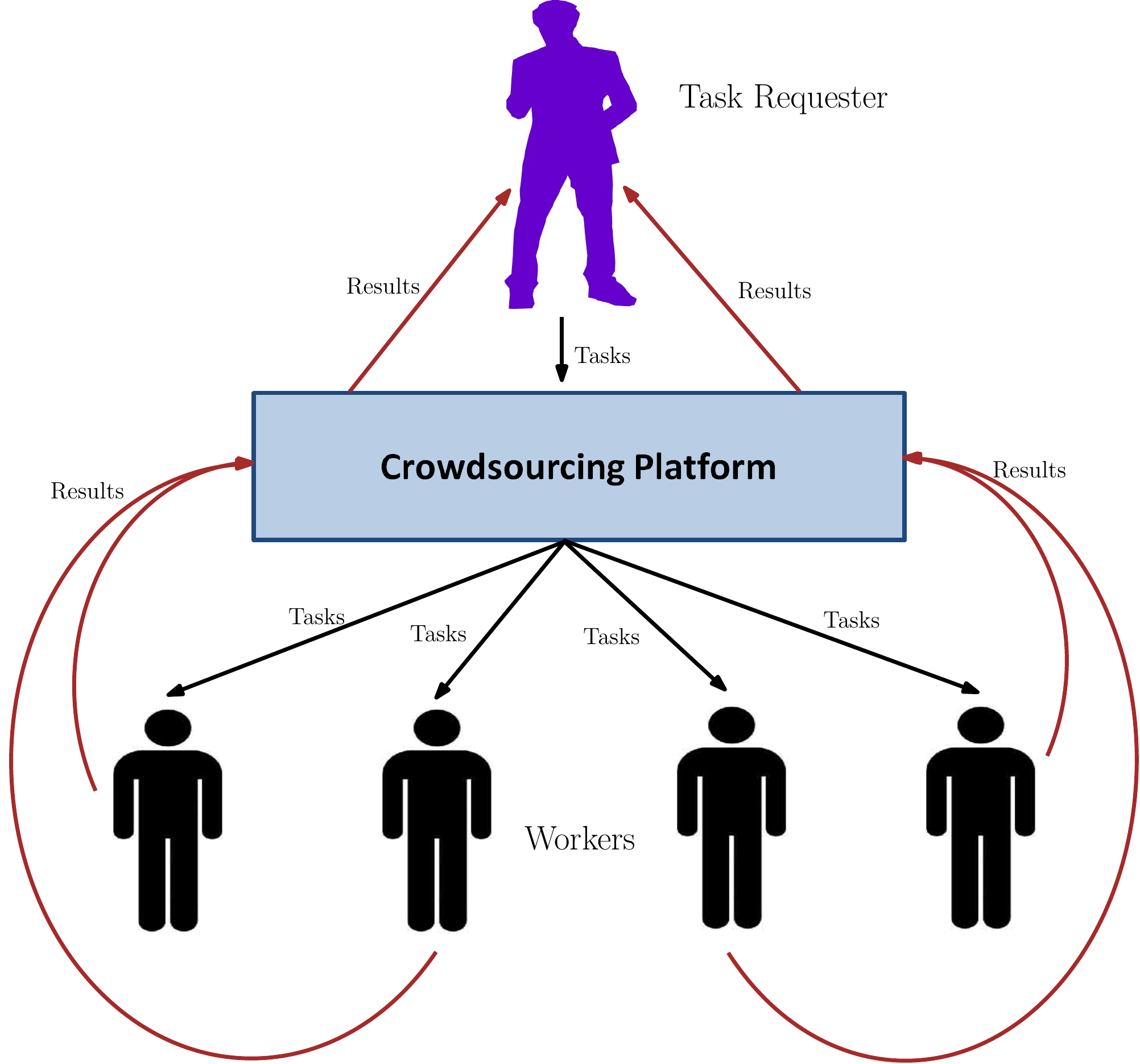}
    \caption{The crowdsouring practice consists of two types of users: task requester and the workers. The task requester distributes tasks to the workers through the platform, and collects the feedbacks from the workers in a reverse way.}
    \label{fig:crowdsouring}
\end{figure*}

Crowdsourcing refers to the distribution of tasks that cannot be easily accomplished in a traditional way to a large group of online workers~\cite{howe2006rise} (Figure \ref{fig:crowdsouring}). The tasks are usually difficult problems or issues that cannot easily be resolved by small groups of users or individuals. Despite its many advantages, crowdsourcing brings increasing risks of information leakage and privacy violation, which limits its development and application potential.

There are two types of users in a crowdsourcing platform: the worker (or the employee) and the task requester (or the employer). The task requester provides incentives and tasks, while the worker performs the tasks to receive the incentives. The interaction between them gives rise to the risks of information leakage and privacy violation, which is either unidirectional or bidirectional. In other words, either the worker or the requester, or both, have the possibility to leak sensitive information or violate the privacy agreement.

We next identify potential privacy leaking risks in three key applications of crowdsourcing: Internet crowdsourcing marketplaces, crowdfunding, and crowdtesting. For each application, we consider the privacy protection issues in the process of sharing practice and survey the existing solutions in the literature.

\subsection{Crowdsourcing Marketplace}\label{sec:internetmarketplace}

An online crowdsourcing marketplace provides a platform for matching the task requesters and the task performers for mutual benefits. Numerous crowdsourcing marketplaces have been developed during the past few years, e.g., the Amazon Mechanical Turk (MTurk)~\cite{kittur2008crowdsourcing}, which enable individuals and business entities to use their own intelligence to perform tasks that are `difficult' for automated computerized programs. Requesters post jobs or work in the form of human intelligence tasks on the MTurk platform, while workers browse the tasks and complete them to earn monetary incentives from the requesters.

Data privacy concerns limit the spreading speed of crowdsourcing because many users refuse to participate in crowdsourcing if personal data cannot be not securely protected. For example, when a requester evaluates the design of a particular artefact, it is likely that the requester desires to prevent exposure of the artefact. Similarly, a testing organization usually requires test takers not to disclose the content of the test. However, unlike a testing organization, which has the power to penalize test takers who violate the confidentiality agreement, the requester does not always have the power or effective methods to penalize workers who leak sensitive data or extract information for other purposes. What makes it worse is that the workers are sometimes unreliable and usually not identifiable. Therefore, it is challenging to protect the privacy of the requesters.

Generally, there are two approaches tackling the privacy protection problem for the requesters. The first solution, which is introduced by Varshney, distorts sensitive data directly using random perturbations to conceal private information~\cite{varshney2012privacy}. A series of extensions were introduced by the same group of researchers for completing the framework based on coding theories \cite{vempaty2014reliable, vempaty2014coding, wang2005distributed}. The coding theory successfully hides the sensitive information from the workers. However, it loses the task performance quality when random perturbations are added to the original data. A mathematical model was used to analyze the tradeoffs between privacy, reliability, and cost, by considering five insight elements: error-correcting codes, reliability, perturbation, decoding and collusion attacks~\cite{varshney2014assuring}.

\begin{figure*}[h!t!b!]
  \centering
  \includegraphics[width=1.5in]{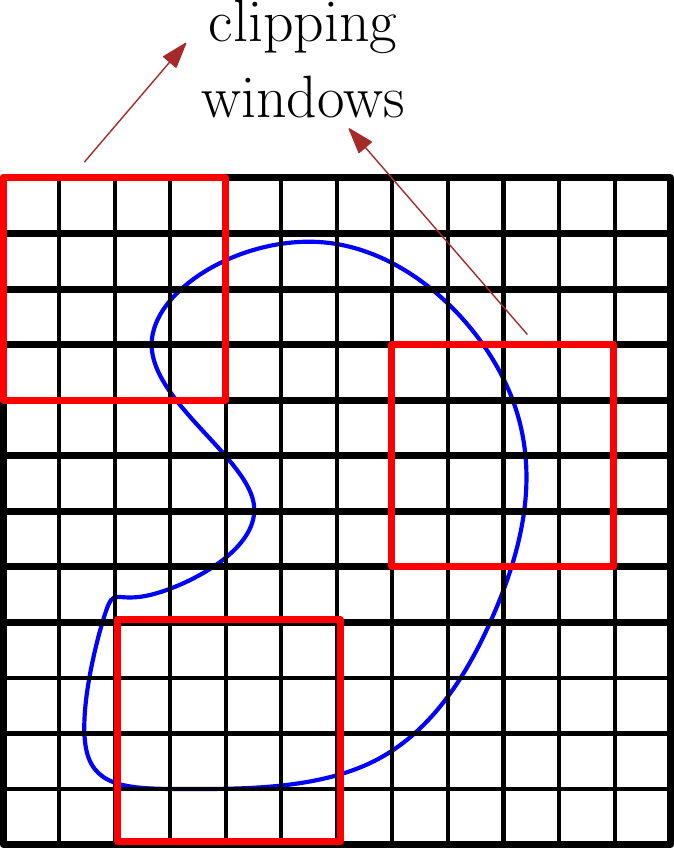}
    \caption{The instance clipping protocol: a task (represented by a 2D shape), is clipped by clipping windows which are marked by red boxes.}
    \label{fig:clipping}
\end{figure*}

The second approach is the instance clipping protocol (ICP), which was introduced by Little and Sun \cite{little2011human} and Chen {\it et al.} \cite{chen2012shreddr}. Kajino {\it et al.} \cite{kajino2014instance} proposed a quantitative analysis framework (QAF) based on the instance clipping protocol. The QAF evaluates the instance privacy-preserving protocols and protects the target privacy, which is defined as contextual information. The instance-privacy preserving protocols preserve instance privacy at the cost of task performance. For instance, in Figure \ref{fig:clipping}, a task (represented by a 2D shape) is clipped by clipping windows which are marked by red boxes. Each worker is only allowed to access one clipping window for his/her task result. The ICP preserves privacy but decreases the quality of the task results. Similar to Varshney's work, there is a tradeoff between privacy preservation and task quality. The instance clipping protocol clips an instance by a moving window, which preserves the data privacy by limiting the data that each worker acquires.

Celis {\it et al.} \cite{celis2016assignment} improved the clipping protocol by introducing a collusion network. The requested task is partitioned into pieces; and each piece of task is assigned to different individuals with minimal privacy leakage. Moreover, a framework is proposed with three operations: PULL, PUSH and Tug Of War (TOW). PULL and PUSH are two usual operations that represent a worker choosing tasks and a requester choosing workers, respectively. The TOW operation is used as an intermediate layer for information leakage minimization, which captures workers' personal information, such as social networks, financial information, task history and etc. However, information leakage is still possible from the workers' side.

Amor {\it et al.} \cite{amor2016discovering} developed a social relationship management system based on clustering algorithms, named `SocialCrowd', to manage competition and collaboration in crowdsourcing practices. Experimental results showed that the data leakage was effectively prevented using SocialCrowd. Since the first version of SocialCrowd uses global search algorithms, the main concern in Amor {\it et al.}'s work is the computational complexity problem. While a heuristic random search method is used in later versions, it can still be trapped into local extremes in worst case scenarios.

\begin{table*}[h!t!b]\caption{References, main objectives, proposed solutions and important insufficiency of the surveyed works for crowdsourcing marketplace.}
\begin{center}
\begin{tabular}{p{2.8cm}p{0.8cm}p{3.6cm}p{3.6cm}p{3.6cm}}
 \hline \hline  \bf{Reference} & \bf{Year} & \bf{Main objective} & \bf{Proposed solution} & \bf{Important insufficiency}\\
 \hline
  {Varshney {\it et al.} \cite{varshney2012privacy, varshney2014assuring}} & 2012, 2014 & {Studying the tradeoffs between privacy, reliability, and cost} &  An improved coding scheme by considering five insight elements & Unable to solve the collision between privacy and task performance quality\\
  {Kajino {\it et al.} \cite{kajino2014instance}} & 2014 & {Protecting the requesters' privacy defined as contextual information} &  Quantitative analysis framework based on instance clipping protocol & Making tradeoff between task performance and privacy\\
  {Celis {\it et al.} \cite{celis2016assignment}} & 2016 & {Partitioning the task with minimal privacy leaks} & The collusion network & Information leakage from the worker side\\
  {Amor {\it et al.} \cite{amor2016discovering}} & 2016 & {Increasing the privacy awareness} & SocialCrowd & Using heuristic function for optimal solution search, which can be trapped in worst case scenarios\\
 \hline \hline
\end{tabular}
\label{table:branch1}
\end{center}
\end{table*}

Table \ref{table:branch1} lists all the references that we have discussed in this section, including their main objectives, proposed solutions, and weaknesses. In summary, while most recent studies of privacy protection in crowdsourcing marketplaces consider coding schemes or clipping protocols, new technologies, such as SocialCrowd, are proposed to help improve the data security. The common problem for the coding schemes and clipping protocols is that the manipulation of the original data decreases the task performance quality. Moreover, the extra time complexity that is added to the original data transmission and storage process is a notable issue for those efforts on privacy protection. In addition, while traditional works focus on protecting the requesters' data in a fundamental way, other issues are raised for improving users' awareness of privacy leakage during crowdsourcing practices. This will be further discussed in Section \ref{sec:emerging1}.

\subsection{Crowdfunding}
Crowdfunding has undergone fast development recently~\cite{belleflamme2014crowdfunding,mollick2014dynamics}. It enables founders of various ventures to fund their projects by collecting funds or other resources from a large group of individuals through an online platform, such as Kickstarter~\cite{kuppuswamy2014crowdfunding} or Indiegogo \cite{blogindiegogo}. While most works focus on economic aspects of crowdfunding, few address privacy issues~\cite{bradford2012new}. To bridge the gap between privacy concerns and practical use of crowdfunding, in this subsection, we review several existing works on privacy concerns in crowdfunding practices.

In the practice of crowdfunding, a fundraiser (the requester) proposes a project with a plan on an online platform and convinces users or supporters to invest small amounts of money in the project. The modern crowdfunding platforms, such as Indiegogo, allow users to customize their security level and conceal their personal information, such as their name and the amount of their contribution. However, our surveyed works suggest that revealing a certain amount of private information can be helpful in crowdfunding practice. For example, concealing the contribution amount of the prior contributor discourages followers from contributing more to the crowdfunding project \cite{burtch2013empirical}. Moreover, a fundraiser may choose to reveal more of his/her personal information to attract crowdfunders \cite{snyder2016crowdfunding}.

\begin{figure*}[h!t!b!]
  \centering
  \includegraphics[width=5.5in]{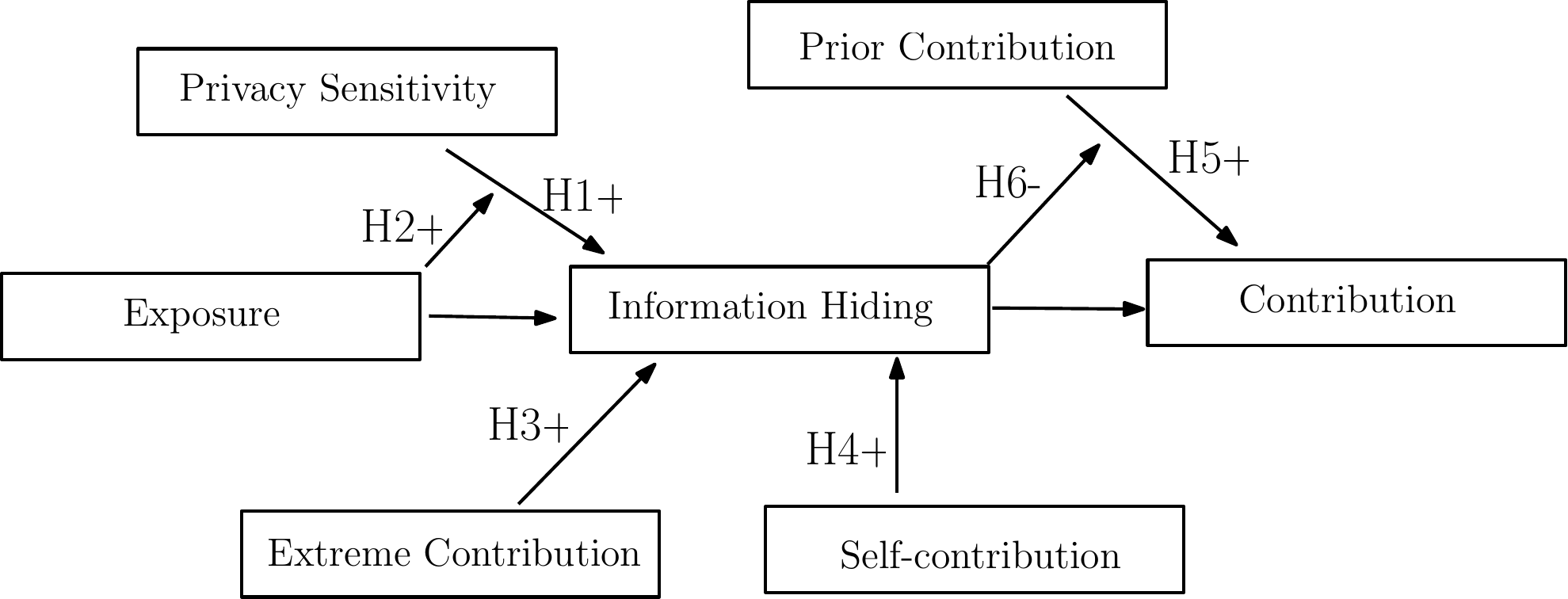}
    \caption{The econometric model proposed by Burtch {\it et al.} \cite{burtch2013empirical, burtch2014experiment}. The likelihood of information hiding and the amount of contribution from crowdfunders are affected by the six hypotheses shown in arrows. The six hypotheses are privacy concern effect (H1), exposure effect (H2), extremity effect (H3), self-contribution effect (H4), anchor effect (H5) and censorship effect (H6). The positive or negative effect is denoted by +/- sign.}
    \label{fig:econometric}
\end{figure*}

Burtch {\it et al.} \cite{burtch2013empirical, burtch2014experiment} conducted a series of experiments on a large-scaled customized crowdfunding platform to test the relationship between the privacy protection level and the results of users' contribution histories. An econometric model was constructed where the dependent variables included the likelihood of information hiding and contribution amount from crowdfunders. The independent variables included the privacy control of the fundraiser's platform, elapsed time of fundraising, and fundraiser's reputation. Six hypotheses were formulated: the privacy concern effect (H1), exposure effect (H2), extremity effect (H3), self-contribution effect (H4), anchor effect (H5) and censorship effect (H6). The econometric model is depicted in Figure \ref{fig:econometric}, where the likelihood of information hiding and the amount of contribution from crowdfunders are affected by the six hypotheses, as shown with arrows. Although the econometric model provided valuable suggestions on privacy protection, it did not consider other factors that influence the crowdfunders' decisions, such as wording, information regulation, transaction mechanism design and etc.

In 2015, Burtch {\it et al.} \cite{burtch2015hidden} conducted another online experiment to study the hidden cost of protecting crowdfunders' privacy by utilizing modern techniques, such as invisible transaction information. Their result indicated that privacy protection increased the net funding in overall, but decreased the contribution amount from each individual. The main insufficiency of \cite{burtch2015hidden} is that all experiments and simulations were conducted in a randomized manner. Moreover, the users were given complete freedom for their fund contributions, which made the experimental result relatively unreliable.

Zheng {\it et al.} \cite{Zheng2016} analyzed the importance of trust management for crowdfunding practices. A research model was constructed for verification of five hypotheses. Experimental results showed that effective trust management techniques significantly improve the fundraising performance. Nevertheless, some important factors, such as funding information and presentation format of funding description, were not considered in the research model, which weakened the reliability of their conclusions.

Kang {\it et al.} \cite{kang2016understanding} introduced a structural equation modeling technique to analyze the true motivations of fundraisers for crowdfundings. Three factors are considered to examine the trustworthiness of a crowdfunding project. The fundraisers' credentials were deeply analyzed by a bootstrapping method that is formed based on historical investment experiences. The main insufficiency of Kang {\it et al.}'s work is that the proposed method was not validated via any cross-sectional surveys.

All reviewed works for privacy issues in crowdfunding practices are listed in Table \ref{table:branch2} Each reviewed work is accompanied by its reference, year, main objective, proposed solution and major insufficiencies. Certain levels of privacy protection, as well as sacrifices, are hidden key factors for successful crowdfunding practices. With a well-established privacy protection protocol, crowdfunders are more willing to contribute because of a safer environment. However, in some situations, a certain degree of acceptable and controllable privacy sacrifice can be helpful for a successful crowdfunding practice. The fundraisers and platforms have to realize that the net funding is directly proportional to their reputations. One open problem is to develop a more sophisticated platform for protecting the funder information. For example, a hierarchical encryption system can be built to serve the basic crowdfunding purposes and allow the fundraisers to select different levels of information sharing with the public for various purposes. Another future research direction is to explore an appropriate degree of fundraisers' privacy disclosure that maximizes the probability of reaching a fundraising goal. Existing works showed that a certain degree of fundraiser's privacy disclosure encourages the funding contributions from users \cite{wen2018information}. However, the most appropriate degree of fundraisers' privacy disclosure remains as an open problem for crowdfunding practices. Existing works showed that a certain degree of fundraiser's privacy disclosure encourages the funding contributions from users~\cite{shen2018information}. Generally speaking, while crowdfunding is a relatively new concept to people in the cyber-enabled world and is directly related to assets, privacy issues are more emerging and are considered one of the most crucial research topics in the development process of the cyber-enabled world.

\begin{table*}[h!t!b]\caption{References, main objectives, proposed solutions and important insufficiency of the surveyed works for crowdfunding.}
\begin{center}
\begin{tabular}{p{2.8cm}p{0.8cm}p{3.6cm}p{3.6cm}p{3.6cm}}
 \hline \hline  \bf{Reference} & \bf{Year} & \bf{Main objective} & \bf{Proposed solution} & \bf{Important insufficiency} \\
  \hline {Burtch {\it et al.} \cite{burtch2013empirical, burtch2014experiment}} &2013, 2014& {Studying the relationship between security and willingness} &  An econometric model & Not taking the full consideration for factors that influence the crowdfunders'  decisions\\
  {Burtch {\it et al.} \cite{burtch2015hidden}} &2015& {Showing the hidden cost of protecting crowdfunders'  privacy utilizing modern techniques} & Online randomized experiments & Experiment users are given complete freedom for their fund contributions\\
  {Zheng {\it et al.} \cite{Zheng2016}} &2016& {Analyzing the importance of trust management} &  A research model based on the elaboration likelihood model & Focusing on the trust management and ignoring other highly influential factors\\
  {Kang {\it et al.} \cite{kang2016understanding}} &2016& {Revealing the fundraiser's true motivation for crowdfunding} &  A structural equation modeling technique & The survey dataset is small in size and limited to only one country\\
 \hline \hline
\end{tabular}
\label{table:branch2}
\end{center}
\end{table*}

\subsection{Crowdtesting}
Crowdtesting employs crowdsourcing technology to employ a large group of testers for software or products testing at low costs \cite{zhang2017crowdsourced}, which is reported to be more reliable, more cost-effective, and faster than traditional user-testing mechanisms~\cite{vukovic2009crowdsourcing, riungu2010research}. One popular crowdtesting platform is well-known as PyBossa~\cite{pybossa2015}, where customized crowdsourcing tasks can be posted, which require human cognition, knowledge or intelligence. The ultimate objectives of a crowdtesting practice include testing usability, acceptability, task performance and the quality of the results.

In a crowdtesting practice, both requesters and workers post crowdsourced data on an online platform, i.e., tasks and results. Part of the crowdsourced data can be privacy related, e.g., the data can include the requester's confidential data and tester's private information. The top priority for privacy preservation in crowdtesting is to protect user privacy in the data collection process. Harkous {\it et al.} \cite{harkous2014c3p} found that users usually had difficulties in accessing the privacy levels of their shared data. A context-aware framework was proposed to identify the privacy risk of shared data on a cloud server. Simulations on synthetic data were performed to show the effectiveness of their method, where data privacy levels were automatically assigned without user interaction. The main limitation of their work is that the proposed system only identifies the risky data items without proposing solutions. Moreover, there is no policy or computational technique proposed in \cite{harkous2014c3p}.

Existing data protection schemes focus on encryption algorithms. Kandappu {\it et al.} \cite{kandappu2013exposing} showed how easily privacy leakage can occur with online survey platforms, such as MTurk and Google Consumer Surveys \cite{mcdonald2012comparing}, which are commonly used in crowdtesting practices. A customized survey platform called Loki was developed to let users choose their preferred security level before proceeding with the online survey. The actual survey results were masked by noises before been evaluated. There are two important insufficiencies in \cite{kandappu2013exposing}. First, the result quality decreased because of the additional noises. Second, there was no guidance for the user to choose the most appropriate security level, which decreases the overall survey quality.

Li {\it et al.} \cite{li2016privacy} explored the privacy issues in crowdsourcing-based site survey systems utilizing WiFi fingerprint-based localization techniques. In a site survey practice, multiple suppliers were required to visit different locations and send back WiFi signals in a crowdsourced manner, which is similar to a crowdtesting practice. The main shortcoming of the work in \cite{li2016privacy} is that the location privacy protection of the suppliers is achieved by encryption and adding noises. The homomorphic encryption can distort the original measurement signal.

Although the crowdtesting service provides an innovative way for services/products to be tested by a large group of testers at low costs, the privacy issues were never well addressed to protect the sensitive information from both the requesters and the testers. Three specific applications of the crowdtesting practices are surveyed in this subsection: shared data protection on the cloud servers \cite{harkous2014c3p}, online surveys \cite{kandappu2013exposing} and indoor site survey practice \cite{li2016privacy}. The objectives, solutions and main insufficiencies are listed in Table \ref{table:branch3}. Almost all reviewed works demonstrate that user privacy can be easily breached by the service providers and platforms in crowdtesting practices. Various techniques were proposed to identify risky shared data and protect those sensitive information pieces. However, encryption or masking of the original data affects the usability of the final testing results, which limits the use of these cyber technologies.

\begin{table*}[h!t!b]\caption{References, main objectives, proposed solutions and important insufficiency of the surveyed works for crowdtesting.}
\begin{center}
\begin{tabular}{p{2.8cm}p{0.8cm}p{3.6cm}p{3.6cm}p{3.6cm}}
 \hline \hline  \bf{Reference} & \bf{Year} & \bf{Main objective} & \bf{Proposed solution} & \bf{Important insufficiency}\\
 \hline
  {Harkous {\it et al.} \cite{harkous2014c3p}} &2014& {Identifying risking data pieces on cloud server} & A context-aware framework based on item response theory (IRT) & Not discussing the way to protect privacy\\
  {Kandappu {\it et al.} \cite{kandappu2013exposing}} &2013& {Allowing users to choose security level for online surveys} &  A customized survey platform called Loki & No guidance for the user to choose appropriate security level\\
  {Li {\it et al.} \cite{li2016privacy}} &2016& {Hiding the location information of the suppliers in indoor site survey practices} & A homomorphic encryption scheme & The original measurement signal was distorted\\
 \hline \hline
\end{tabular}
\label{table:branch3}
\end{center}
\end{table*}

\subsection{Summary and Discussion}

In conclusion, in crowdsourcing practices, there are always three roles in the model: user, requester and platform. On one hand, the requester has the responsibility to protect workers' privacy. On the other hand, the requester designs mechanisms or protocols that discourage workers from leaking sensitive data of the tasks; and the workers are responsible for following the privacy agreements of tasks. The platform serves as a mediator that protects the privacy of both parties. Both the task requester and the users must understand that there are always tradeoffs between privacy and interests (e.g., incentives, task quality, funds). Both entities must sacrifice part of their privacy to enjoy a quality crowdsourcing practice. For example, in the crowdfunding practice, a reliable platform protects the privacy from both the users' and requesters' perspective, which increases the trust between both parties and further increases the chance of successfulness of the crowdfunding campaign \cite{wen2018information, liang2018why}.

While most of the works that are surveyed in this section focus on cyber technology development on the platform for protecting the privacy of both the task requesters and workers, some policy/regulation works are mentioned as supplementary materials. Although the business models of these three crowdsourcing practice branches are different, raising the privacy protection level is always helpful to both the workers and task requesters in achieving their goals.

In general, on a crowdsourcing platform, users should be allowed to retrieve information from the database of a sharing service provider while the queries are maintained privately. In addition, to increase the security level of data protection for users, data de-identification methods are available in most cases~\cite{graham1977dynamic,uzuner2007evaluating,gilbert2001identification,el2009globally}. Traditional methods, such as $k-$anonymity, $l-$diversity models, etc., can also be used to avoid linkage attacks ~\cite{samarati2001protecting, samarati1998generalizing,bayardo2005data}.

\section{Privacy Issues and Solutions in Collaborative Consumption Practices}\label{sec:collaborative}

\begin{figure*}[h!t!b!]
  \centering
  \includegraphics[width=4.5in]{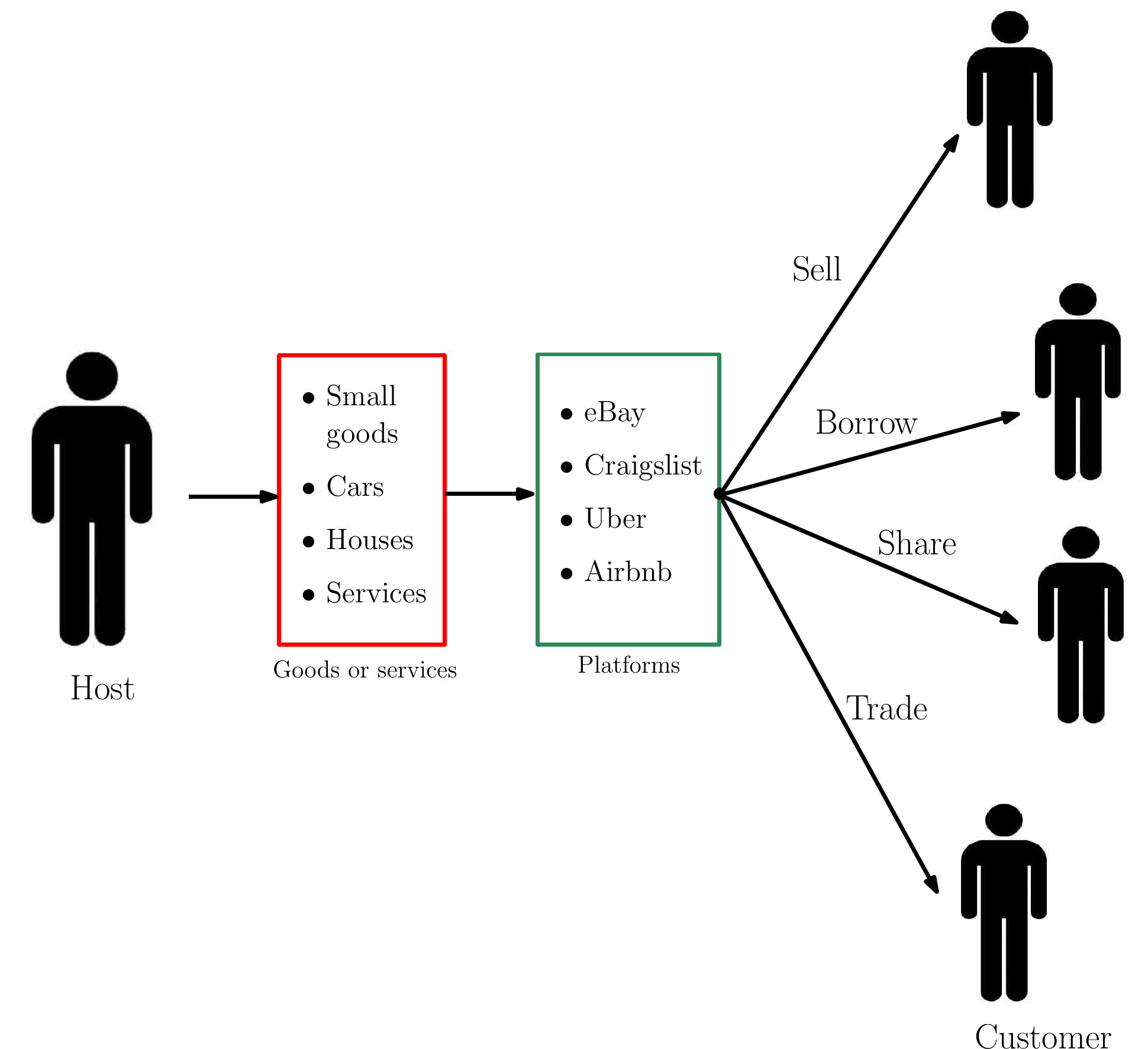}
    \caption{A typical demonstration of collaborative consumption: hosts provide shared goods or services through a platform to the customers. The sharing methods include selling, borrowing, trading and sharing; the typical examples of collaborative consumption platform include eBay, Craigslist as well as Uber (ridesharing) and Airbnb (homesharing).}
    \label{fig:collaborate}
\end{figure*}

Unlike crowdsourcing-based sharing services, which combine the power of a large group of individuals to perform tasks, collaborative consumption allows individuals to access goods or services through P2P sharing that is coordinated by online platforms~\cite{belk2014you, botsman2011s}. In collaborative consumption practices, hosts provide shared goods or services through a collaborative consumption platform to customers. The sharing methods can be selling, borrowing, trading and sharing. Typical examples of collaborative consumption platforms include eBay (collaborative Online shopping), Uber (ridesharing) and Airbnb (homesharing) (Figure \ref{fig:collaborate}). Collaborative consumption has many benefits, such as greenhouse gas emissions reduction, cost saving, unaffordable goods access and decentralization~\cite{hamari2015sharing, botsman2011s}. Although collaborative consumption has many advantages, it suffers from privacy concerns. In this section, we review problems and solutions related to privacy issues in collaborative consumption.

\subsection{Collaborative Online Shopping}

Online shopping is probably the first successful model in which cyber technology has changed our living world. In the first stage of online shopping development, people found that it was more convenient and economical to purchase goods over Internet. In the process of cyber technology development, the concept of collaborative consumption was gradually embedded into the online shopping experience. People started to sell small items, trade services, share cars and borrow items through online shopping websites \cite{botsman2011s}.

On the other hand, online shopping websites have received many criticisms due to their notorious privacy policies despite their popularity~\cite{wright2001controlling, light2013sure, bowie2006privacy, valentine2000privacy}. Although it is illegal to reveal user information to third parties without user consent, online platforms are not subject to a penalty for analyzing user data. These platforms rely on third-party organizations for data analysis, which deteriorate customers' privacy. The privacy policy terms are supposed to be accepted by customers without negotiations, which are in some sense unfair to the customers. Except for limited government regulations, these marketplaces are self-regulated or autonomous, which makes it difficult to protect consumer's privacy. Moreover, these platforms suffer from data leakage due to cyber attacks or intrusion. These factors contribute to the vulnerability of consumers' privacy.

Miyazaki and Fernandez \cite{miyazaki2001consumer} surveyed about online shopping fears on a set of U.S. Internet users from different age groups, economical classes and educational backgrounds. The survey results indicated that the untrusted security system is the largest fear of the customers. Malhotra {\it et al.} \cite{malhotra2004internet} systematically analyzed Internet users' information privacy concerns (IUIPCs) through two separate surveys of 742 household respondents. They designed a theoretical framework for studying IUIPCs and proposed a causal model that predicts the reaction of online customers to privacy threats from shopping websites. Tsai {\it et al.} \cite{tsai2011effect} studied how the privacy concerns of customers affected their decisions in the online shopping process. They conducted an experiment to test the shopping decisions that were made by customers after displaying their personal information on the shopping websites. Their results demonstrate the customers' willingness to pay a premium for extra privacy protection (from a more expensive shopping website). All of the above mentioned works reveal the fact that the privacy concern is the main fear in online shopping experiences. However, these works do not present a deep analysis on how to build privacy protection trust between online shopping websites and customers using regulation policies or cyber technology.

\begin{figure*}[h!t!b!]
  \centering
  \includegraphics[width=4.5in]{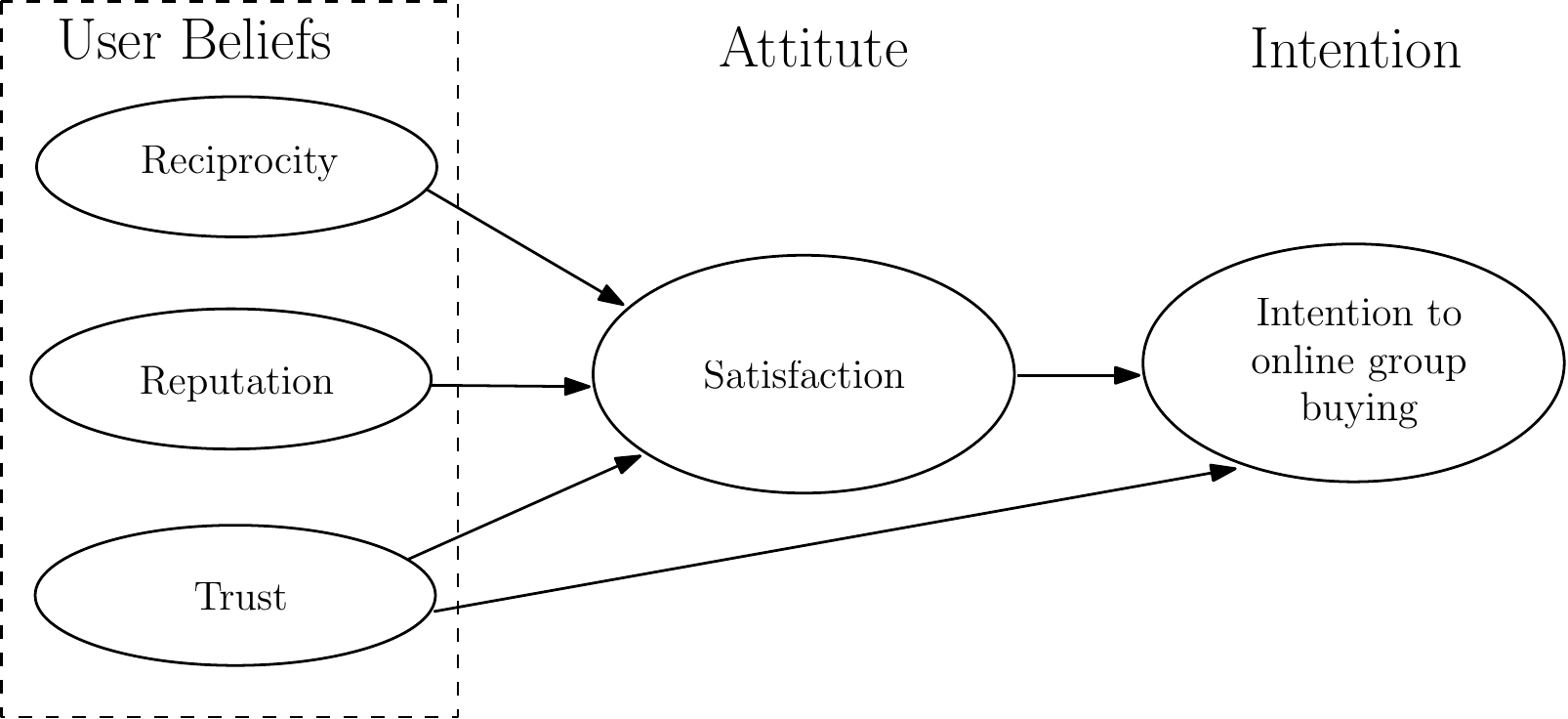}
    \caption{The research model proposed by Shiau and Luo \cite{shiau2012factors}, showing the relationship between consumer satisfaction, intension of online group buying and user beliefs.}
    \label{fig:shiau}
\end{figure*}

Shiau and Luo~\cite{shiau2012factors} built a research model using partial least squares (PLS) method to analyze the relationship between consumer satisfaction, intention of online group buying and user beliefs (Figure \ref{fig:shiau}). The PLS results show that consumer satisfaction highly depends on trust, followed by reciprocity. It is the first work to draw an overall picture of the different factors that affect the online shopping decisions. Moreover, it is also the first work to clearly identify privacy concern as the first priority for online shopping security. Following Shiau and Luo's work, Bergstr{\"o}m \cite{bergstrom2015online} built an analytic system with different groups of people concerning various privacy issues in online shopping experiences. Both the customers and the privacy concerns were partitioned into different dimensions to interpret the links between socialization, Internet experience, trust, politics, and security understanding. Their analysis result clearly indicated that the trust is the major concern of people who worry about the misuse of personal data. Although these research models go one step further than the simple survey results, they still do not provide a clear solution for protecting the customers' privacy in online collaborative shopping practices.

Preibusch {\it et al.} \cite{preibusch2016shopping} studied and reported a concrete example of privacy leakage in online shopping practices. They performed online tracking and found that online shopping websites send unnecessary personal information to payment providers, such as Paypal. Therefore, there is an on-going risk for customers who shop online. The most effective method for changing this situation is to facilitate relevant legislation. However, the lack of government regulation of online shopping websites exists globally. Moreover, it remains unclear what rules can be added and how they can be enforced. Although there are existing regulations (Directive 95/46/EC by the European Union \cite{poullet2006eu} and USA Patriot Act \cite{kerr2003internet}), existing studies have shown that those regulations are usually ignored due to insufficient government monitoring.

One solution to protect users' privacy in collaborative online shopping practices is to install third-party privacy protection software in the web browser. Available software on Internet includes the Tor Browser \cite{macrina2015tor}, the Privacy Bird \cite{vu2016user} and the Platform for Privacy Preferences \cite{perera2015end}. These third-party software programs or plugins identify untrusted shopping websites and mask personal information for the customers. However, third-party software is usually not formally authorized or registered by the government, which potentially raises other concerns of privacy leakage.

\begin{figure*}[h!t!b!]
  \centering
  \includegraphics[width=4.5in]{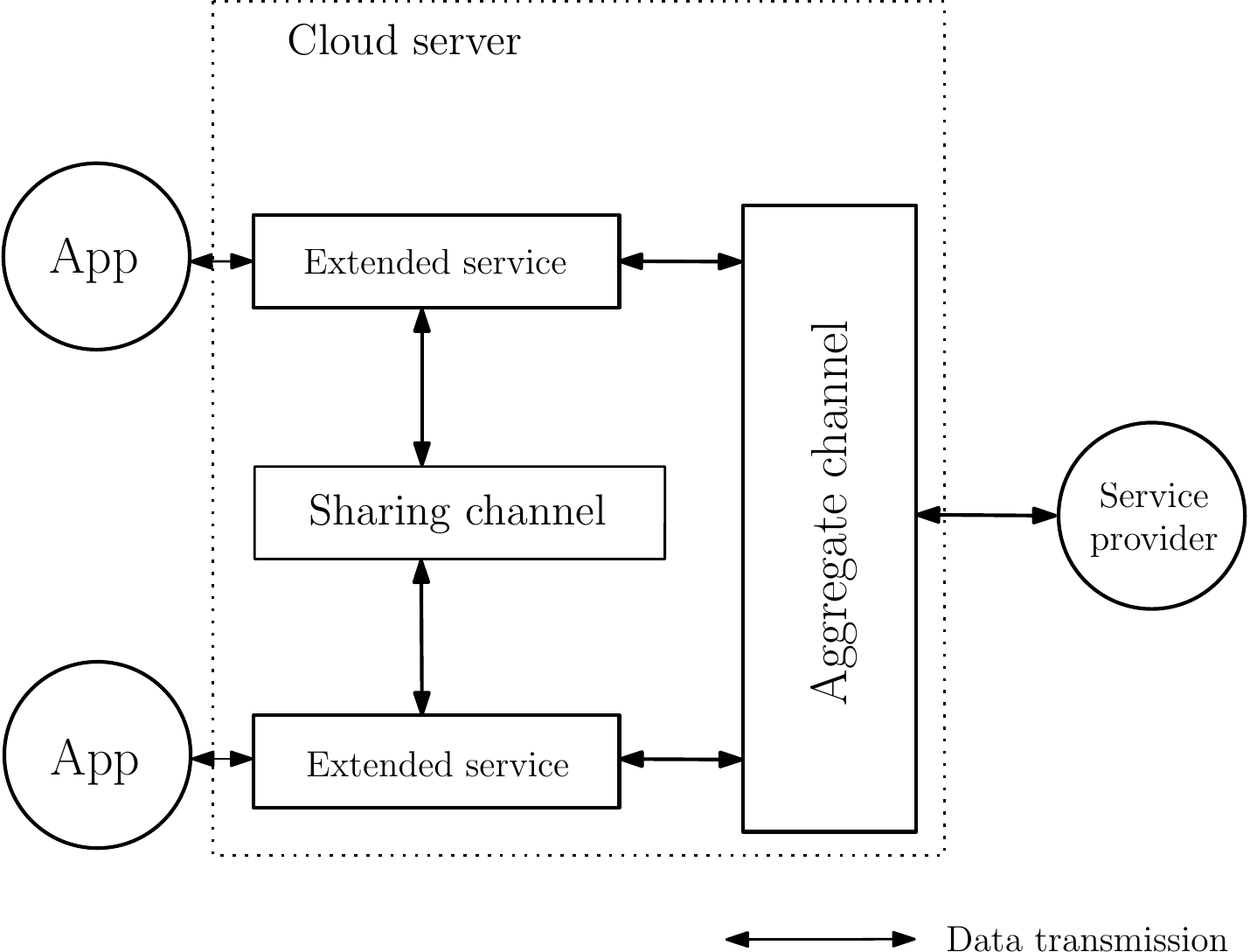}
    \caption{The internal structure of $\pi$-box: it extends the user apps on cloud server. Two separated channels were designed, which were the sharing channel controlling the data transmission between users and the aggregate channel controlling the data transmission from users to the service provider.}
    \label{fig:pibox}
\end{figure*}

Lee {\it et al.} \cite{lee2013pibox} proposed a $\pi$-box mobile app to control the sensitive data transmission between different users and from users to service providers. The $\pi$-box extends the user apps and was built based on the cloud services that were supplied by large companies, such as Google. Two separate channels were designed: the sharing channel, which controls the data transmission between users and the aggregate channel; and the aggregate channel, which controls the data transmission from users to the service provider. The structure of $\pi$-box is illustrated in Figure \ref{fig:pibox}. All channels are internally monitored by a centralized system. The limitation of the proposed $\pi$-box is that it does not universally apply to any app in market. According to a user survey conducted by Lee {\it et al.} \cite{lee2013pibox}, only 48\% of paid apps support $\pi$-box, which limits its usage on privacy protection.

Kokolakis \cite{kokolakis2017privacy} studied the conflict between the customer's high demand for privacy protection and the customer's willingness to sacrifice privacy for the exchange of goods or services in the online shopping practice. Kokolakis concluded that this inconsistency represents a collision between a customer's attitude and behaviour, which is known as the privacy paradox \cite{Norberg2007The}. A large volume of works was surveyed to justify the existence of the privacy paradox; however, most of them are surveys or experimental works that do not involve theoretical model.

Bilge and Polat \cite{bilge2013comparison} introduced a method for improving the online shopping experience by collecting customers' personal information, such as ratings and comments for a particular service, in a privacy-preserving manner. A number of clustering methods were integrated into the collaborative filtering service. The system filtered out customized information by training on encrypted user data using clustering methods. The main insufficiency of the work in \cite{bilge2013comparison} is that, due to the encryption of the users' data, the recommendation error rates increased. In addition, the clustering methods introduced extra computational costs to the recommendation system.

The reviewed works, which are listed in Table \ref{table:branch4}, identified two privacy threats in collaborative online shopping practice. The first threat comes from the service provider, where unreliable platforms may misuse customers' data for marketing analysis. This threat can be prevented by refining government regulations \cite{preibusch2016shopping}, masking customers' data before sending them out \cite{bilge2013comparison} or separating communication channels on the cloud server \cite{lee2013pibox}. Other possible solutions to prevent such malicious behaviours include utilizing trusted computing \cite{Santos2009Towards} or building services based on a trusted provider \cite{Paladi2017Providing, Paladi2014Domain}. The second threat comes from the customer side, where most customers realize that they must sacrifice a certain degree of privacy to enjoy the collaborative shopping experience \cite{kokolakis2017privacy}. It is difficult for them to choose a trustworthy service provider, products \cite{bergstrom2015online}, and most importantly, the kinds of permissions to grant \cite{ismail2015crowdsourced}. The second threat can be alleviated by increasing the overall privacy awareness of the users, which will be extensively discussed in Section \ref{sec:emerging1}.

\begin{table*}[h!t!b]\caption{References, main objectives, proposed solutions and important insufficiency of the surveyed works for collaborative Online Shopping.}
\begin{center}
\begin{tabular}{p{2.8cm}p{0.8cm}p{3.6cm}p{3.6cm}p{3.6cm}}
 \hline \hline  \bf{Reference} & \bf{Year} & \bf{Main objective} & \bf{Proposed solution} & \bf{Important insufficiency}\\
  \hline {Miyazaki and Fernandez  \cite{miyazaki2001consumer}}, {Malhotra {\it et al.} \cite{malhotra2004internet}}, {Tsai {\it et al.} \cite{tsai2011effect}} &2001, 2004, 2011& {Pointing out the biggest fear for online shopping experiences} & Surveys on Internet users & Lacking a deep analysis to build the privacy protection trust between the online shopping websites and the customers\\
  {Shiau and Luo \cite{shiau2012factors}, Bergstr{\"o}m \cite{bergstrom2015online}} &2012, 2015& {Learning the largest privacy concern in online shopping practices} &  Drawing the overall online shopping fears relationships by research models & No clear solution for protecting the online customers' privacy\\
  {Preibusch {\it et al.} \cite{preibusch2016shopping}} &2016& {Pointing out the need of raising government regularization for online shopping globally} & A concrete example of the privacy leakage in online shopping practices & What rules to be added and how to add are two big questions\\
  {Lee {\it et al.} \cite{lee2013pibox}} &2013& {Separating the data transmission from user to user and from user to service provider} & A mobile app called $\pi$-box & Not supporting all paid apps\\
  {Kokolakis \cite{kokolakis2017privacy}} &2017& {Revisiting the conflict known as `privacy paradox'} & A survey covering related existing works & No theoretical model was discussed\\
  {Bilge and Polat \cite{bilge2013comparison}} &2013& {Protecting privacy in user information collection process for a recommend system} & Masking sensitive data and using clustering methods for data analysis & Losing analysis accuracy\\
 \hline \hline
\end{tabular}
\label{table:branch4}
\end{center}
\end{table*}

\subsection{Ridesharing}

Real-time ridesharing or dynamic carpooling is a transportation service that allows commuters to share rides on very short notice through mobile apps~\cite{chan2012ridesharing, agatz2011dynamic, shen2015managing, shen2016online, shen2017regulating}. Successful ridesharing platforms, such as Uber, are available in most major cities in the world. When a user needs a ride, he/she may simply use a mobile app to request a ride by entering the destination. The app provides the estimated cost and assigns a driver to the passenger. The payment is generally made with the credit card or other digital payment methods that are associated with his/her account. In the end, both the passenger and the driver will rate each other.

It is well known that the mobile apps can track the customers' location information and travel information for better service quality. The driver has to access to the rider's travel information, such as riders' names, trip starting points and destinations to provide services. Under current privacy policies, riders have to share part of their private information to receive ridesharing services. The platforms have limited regulatory power over the drivers because the drivers are contractors rather than employees of the ridesharing companies. Moreover, drivers' names and license plate information are also subject to disclosure. Concerns have been raised about the internal misuse of user data within the ridesharing companies. For instance, staffs in the ridesharing companies have the access to data for tracking the movements of customers. Taking Uber as an example, in its user agreement terms, it is clearly stated that user information, such as the geo-location, is recorded and internally used by the company for research development purposes. However, the purposes of internal research are not defined explicitly. Customers worry about how their private data is used. Additionally, Uber can access, use, preserve, transfer and disclose user information to prevent, discover or investigate violations of the privacy policy or the user agreements as determined necessary or appropriate. However, customers do not know what information is necessary or appropriate.

Location privacy has been studied extensively in recent decades because of the pervasiveness of geo-location related software and mobile apps~\cite{beresford2003location, mitrokotsa2014location, shahandashti2015reconciling, bettini2005protecting, barkhuus2003location}. While location-aware applications track customers' location or other data online, they generate a huge amount of potentially sensitive data. The privacy of location data depends on the regulation of data access. It is neither necessary nor possible to forbid all accesses because the systems must access the data for analysis purposes. Moreover, access permissions should be given to authorized persons and should never be exposed to others. In other words, the data and the access should be tightly controlled and data should be accessed only with legal authorization~\cite{beresford2003location}.

Kido {\it et al.}~\cite{kido2005protection} proposed one of the first techniques for concealing the actual locations of customers in location-based services, including ridesharing practices. When a user sends an inquiry to the server, he/she sends his/her actual location, together with two false positions called `dummies.' The dummy nodes in the tracking system are carefully generated such that an observer cannot easily identify the actual location of the user; however, the location-based server (LBS) can find the difference through optimized algorithms with external information such as road navigation service (RNS) data. The obvious shortcoming of Kido {\it et al.}'s work is that the real location is not completely concealed (by using dummies). There is still a chance that the observer will identify the actual location.

Yao {\it et al.} \cite{yao2010clustering} provided an effective encryption service for ridesharing customers using the clustering $k$-anonymity (CK) scheme \cite{samarati2001protecting}. The CK scheme encrypts the user location information by utilizing a cloaked spatial-temporal boundary (CSTB) that involves $K$ users. The spatial and temporal constraints, which determine the resolution of the encryption, can be customized by users. However, the use of CSTB decreases location information resolution, and consequently, degrades the service quality of ridesharing.

Pan and Meng \cite{pan2013preserving} extended Yao {\it et al.}'s work using a $p$-anti-conspiration model for location privacy protection. Various techniques were introduced, including methods that provide LBS without knowing the actual locations of the customers. It is a large advancement for the ridesharing companies in protecting the user locations. A follow-up work done by the same group of authors in \cite{pan2016protecting} showed that the approach proposed in \cite{pan2013preserving} lacks protection on sensitive information during the data transmission process.

Jagwani and Kaushik~\cite{jagwani2012defending} intended to prevent location information leaks using the concept of Zero knowledge proof (ZKP). The construction process of the authentication scheme based on ZKP was introduced; and the possible applications of ZKP in the location-based service domain were discussed. The main shortcoming of the ZKP approach is that an authentication scheme is always required to coordinate between customers and hosts.

Gao {\it et al.} \cite{gao2013trpf} introduced trajectory privacy in the ridesharing practices. The trajectory privacy contains spatial-temporal information, which is an important addition to the location privacy protection scheme. In their study, they proposed a mixed-zone graph model to protect the trajectory privacy. The actual implementation relies on a third party middleware, where the actual location information leakage exists.

In recent years, online social networks or geosocial information have started to be used in ridesharing services. It is preferable to use a friend's car rather than stranger's. Based on this motivation, Elbery {\it et al.} \cite{elbery2013carpooling} proposed a social Vehicular Ad-Hoc Network (S-VANET) carpooling recommendation system. They embedded friendship locality, preference locality, and travel locality information into the ridesharing recommendation system, which requires a large amount of privacy information from both the requester and his/her friends.

Ni {\it et al.} \cite{ni2016ama} suggested that customers' true identities can be hidden by incorporating an anonymous mutual authentication (AMA) protocol into the carpooling recommendation system. A real-time navigation system is proposed for concealing the drivers' privacy \cite{ni2016privacy}. One important feature of their application system is the false information traceability, where the trusted third party authority can trace incorrect information, either from a user or a driver. The main limitation of their work is that a trusted third-party middleware is still required.

A{\"\i}vodji {\it et al.} \cite{aivodji2016meeting} proposed a privacy-preserving local computational method for determining the meeting point of a driver and a rider in a ridesharing system, which does not require third-party middleware. Multimodal routing algorithms are used to compute a mutually interested meeting point for both the driver and rider. However, the current developed system was only designed to accommodate one driver and one rider. A more sophisticated system that can include multiple drivers and riders for ridesharing practices is left as a future work.

Shokri {\it et al.} \cite{shokri2016privacy} concluded that the current location privacy protection approaches can be concluded on three trends, which are perturbing the actual location, tracing the perturbed location, and evaluating the privacy-preserving methods. While most existing works only focus on encrypting the customer's current location, strategies were employed by attackers to trace down the actual location of the customer. Useful private information pieces, such as recently visited locations, frequently visited places and nearby landmark buildings, become potential clues for the attackers in estimating the current location of the customer. In \cite{shokri2016privacy}, a comprehensive Bayesian security game is designed to simulate various cases in which a strategic attacker traces the actual location of a customer. Four different scenarios were studied. However, it was difficult to predict the intelligence level of the attacker; and the whole simulation system is too complex in most of the real-world scenarios.

Vergara-Laurens {\it et al.} \cite{vergara2017privacy} categorized privacy preserving systems into approaches for two processes: the tasking process, where tasking devices (such as mobile phones) collect data in certain areas; and the reporting process, where distributed devices report sensed data to the platform. Both processes exist in ridesharing practices. Three open problems were raised for crowdsensing (CS) researchers in the field of location privacy preservation, which are 1) privacy-preserving mechanisms for tasking processing, 2) privacy-preserving mechanisms for reporting process and 3) selecting the most appropriate privacy-preserving mechanism.

Wang {\it et al.} \cite{wang2018truthful} proposed a two-stage auction algorithm taking both trust degree and privacy sensibility into consideration for mobile crowdsourcing systems, such as ride-sharing practices. The $k$-anonymity scheme is integrated with $\epsilon$-differential scheme to add Gaussian white noise to the actual locations of users. The proposed scheme was proven to be trustful and can inspire more users to participate in the mobile croudsourcing systems. Insufficiency exists while the added Gaussian white noise increases the computational complexity and consequently weakens the service quality for mobile crowdsourcing systems.

\begin{table*}[h!t!b]\caption{References, main objectives, proposed solutions and important insufficiency of the surveyed works for ridesharing.}
\begin{center}
\begin{tabular}{p{2.8cm}p{0.8cm}p{3.6cm}p{3.6cm}p{3.6cm}}
 \hline \hline  \bf{Reference} & \bf{Year} & \bf{Main objective} & \bf{Proposed solution} & \bf{Important insufficiency}\\
 \hline {Kido {\it et al.}~\cite{kido2005protection}} &2005& {Protecting location privacy using dummies} & An anonymous communication technique & The actual location is not completely concealed\\
  {Yao {\it et al.} \cite{yao2010clustering}} &2010& {Encrypting the user location information} &  Clustering K-anonymity (CK) scheme & Decreasing location information resolution and degrading the QoS\\
  {Pan and Meng \cite{pan2013preserving}} &2013& {Providing location-based services without knowing the exact location} &  The $p$-anti-conspiration privacy model & lacking protecting sensitive information\\
  {Jagwani and Kaushik \cite{jagwani2012defending}} &2012& {Removing the dependency of using third party software} & Zero knowledge proof & An authentication scheme is required \\
  {Gao {\it et al.} \cite{gao2013trpf}} &2013& {Protecting the trajectory privacy} &  A trajectory privacy-preserving framework & The exact location must be revealed to a third party middleware\\
  {Ni {\it et al.} \cite{ni2016ama, ni2016privacy}} &2016& {Concealing both customers and drivers' sensitive information} &  An anonymous mutual authentication (AMA) protocol & A trusted third party middleware is required\\
  {A{\"\i}vodji {\it et al.} \cite{aivodji2016meeting}} &2016& {Computing the mutually interested meeting point} & Multimodal routing algorithms & A more complicated system involving multiple drivers and riders are left for future exploration\\
  {Shokri {\it et al.} \cite{shokri2016privacy}} &2016& {Considering strategic attackers for customer's location privacy} & A comprehensive Bayesian security game & The complexity is not necessary for most of the real-world scenarios\\
  {Vergara-Laurens {\it et al.} \cite{vergara2017privacy}} &2017& {Surveying privacy-preserving mechanisms} & A survey of all existing works on location privacy preservation & Only location privacy is heavily surveyed\\
  {Wang {\it et al.} \cite{wang2018truthful}} &2018& {Inspiring more users to participate in the mobile croudsourcing systems} & Integrating $k$-anonymity scheme with $\epsilon$-differential scheme & The addition of Gaussian white noise weakens the crowdsourcing service quality\\
 \hline \hline
\end{tabular}
\label{table:branch5}
\end{center}
\end{table*}

All reviewed papers are summarized in Table \ref{table:branch5}. Similar to other sharing services, customers realize that a certain degree of their privacy must be sacrificed to enjoy better service quality. Taking the Uber service as an example, the platform (Uber app) usually records the customers' private information, including current location, destination, phone number, recent trips and so on, to serve them better. However, the customers sacrifice their privacy to enjoy the Uber service. The conflict between the disclosure of private information and the service quality becomes more obvious in the ridesharing practices, which is also mentioned in most of the surveyed works, such as \cite{gao2013trpf, ni2016privacy, aivodji2016meeting, shokri2016privacy}.

Compared to other fields of sharing service, ridesharing is a relatively new technology. Few regulations have been established in this area; and most privacy concern solutions are on technical aspect. Despises the variety of technologies proposed by the existing works, only location privacy is extensively discussed. Ridesharing services include direct interpersonal interactions (IPIs), e.g., the conversation between the rider and the driver when they are travelling \cite{Vertommen1980The, Andersen2002The, cho2011interpersonal}. Computerized technologies, which are designed to be embedded in the online platform, can be helpless in IPI; and physical privacy concerns exist at this stage \cite{christin2016privacy}. Physical privacy concerns, which were first defined by Belk, occur when the driver or passage's personal space is invaded, where we refer to the remaining privacy concerns as online privacy concerns \cite{belk1988possessions, belk2010sharing}. For future works in this field, we would like to note that physical privacy protections for both the riders and drivers are demanded in the ridesharing practice.

\subsection{Homesharing}

Homesharing is a business model that connects hosts and travellers through an online marketplace platform and enables transactions without the platform owning any rooms itself. It does not provide the rental services directly. Instead, it matches hosts who have extra rooms for rent and travellers who need a room for stay~\cite{panda2015emergence,gaikar2013first}. One of the most famous homesharing platforms is Airbnb~\cite{jefferson2014airbnb}.

The face-to-face e-commerce model makes the physical privacy issue more serious for homesharing practices compared with online sharing model. The host and traveller usually meet each other before a deal was made and both of them have the possibility to reveal the privacy of each other to the public. For example, a host may install a hidden camera in an Airbnb room to monitor travellers. A traveller may take pictures to reveal the details of the room or other parts of the house to the public. The online platform records sensitive information of both the hosts (e.g., names, travel plans) and the travellers (e.g., names, home locations).

Kamal \cite{kamal2016trust} realized that the largest inhibitor of homesharing services is the fear of privacy disclosure. They argued that additional background checks are always necessary for participants in homesharing activities, with the possibility of additional security measures, such as certificates and safety insurance. However, we would like to point out that the cost comparison between the additional security checks and the actual accommodation is not discussed in \cite{kamal2016trust}.

Morosan and DeFranco \cite{morosan2015disclosing} determined the level of willingness of travellers to disclose their personal information to hotel apps. An extended version of the privacy calculus model was adopted. The experimental results indicated that personal information disclosure was indeed helpful for the hotel business, i.e., to choose the best customers. But the willingness of disclosing such information was related to privacy concerns, trust, emotions and etc. The main insufficiency of this work is that the study data was collected from U.S. customers who were involved in a relatively safe environment with reliable network security, regulations and hotels. The experimental results may not be applicable to third-world countries.

Ert {\it et al.} \cite{ert2015trust} designed an experiment that used mixed-logit analysis to determine the relationship between the posting of a host's photo in the advertisement and the booking likelihood. The results show that both the trustworthiness and attractiveness of the host's photo increase the likelihood of the house being booked. Similar to crowdfunding and crowdtesting, an appropriate degree of private information disclosure from the hosts's side increases the probability of success for the entire practice/business. However, on the other hand, the leakage of the hosts' privacy, including posting of the host's photo and identity information, is another issue in homesharing practices, which is deeply discussed by Hooshmand \cite{hooshmand2015risks}.

Lutz {\it et al.} \cite{lutz2017role} explicitly divided the privacy concerns into physical privacy concerns (e.g., physical damages of private assets) and online privacy concerns (e.g., personal identity leakage). They conducted a survey on MTurk involving 389 participants; and most of them were hosts on Airbnb. The survey results showed that physical privacy concerns are more crucial than online privacy concerns in the homesharing business. The main shortcoming of their work is that the survey is limited to Airbnb hosts and does not include any customers. Thus, the survey results may be biased towards the hosts' preferences.

We list all reviewed works for security concerns of homesharing in Table \ref{table:branch6}. Similar to crowdsourcing practices, certain degrees of private information disclosure from the hosts side positively influence the trust from the customers side and consequently attract more customers. Moreover, compared to other sharing services, homesharing involves more interpersonal interactions. Concerns about physical privacy are heavily studied in this field. Most of our surveyed works agreed that the hosts are more concerned about their privacy leakage than the travellers. Future studies can focus more on the development of privacy protection schemes for hosts. In the current stage of homesharing, while it is unlikely to solve the privacy issue with a single method, it is quite possible to provide a general privacy-preserving environment for both hosts and travellers through the joint efforts of hosts, travellers, platforms, and governments.

\begin{table*}[h!t!b]\caption{References, main objectives, proposed solutions and important insufficiency of the surveyed works for homesharing.}
\begin{center}
\begin{tabular}{p{2.8cm}p{0.8cm}p{3.6cm}p{3.6cm}p{3.6cm}}
 \hline \hline  \bf{Reference} & \bf{Year} & \bf{Main objective} & \bf{Proposed solution} & \bf{Important insufficiency}\\
  \hline
  {Kamal \cite{kamal2016trust}} &2016& {Building trust in homesharing practices} &  Proposing additional security checks & Not discussing the cost of additional security measurements\\
   {Morosan and DeFranco \cite{morosan2015disclosing}} &2015& {Checking the willingness of hotel customers to disclose personal information} & An extended version of the privacy calculus model & Collecting data only based on U.S. customers\\
  {Ert {\it et al.} \cite{ert2015trust}} &2015& {Showing the relationship of posting a host's photo and the booking likelihood} & Mixed-logit analysis & The relationship between trust and privacy is of interest but not discussed\\
  {Lutz {\it et al.} \cite{lutz2017role}} &2017& {Investigating the impact of physical privacy concerns to homesharing} & A survey on MTurk & Only hosts were surveyed\\
 \hline \hline
\end{tabular}
\label{table:branch6}
\end{center}
\end{table*}

\subsection{Summary and Discussion}

Collaborative consumption collects extra information resources and distributes them to people who do not have access to them. Users are subject to privacy leakage due to the exchange of data and improper use of user data by internal staffs or the platforms. Similar to crowdsourcing practice, both hosts and customers must understand that certain degrees of their privacy have to be sacrificed for better service quality. The most appropriate degrees for private information disclosure from both hosts and customers side are left as open problems to maximize the service quality and net profit. While it is difficult to provide an absolute privacy-safe environment without sacrificing service quality, it is possible to increase the protection levels of privacy through a joint effort of all participants, platforms and governments \cite{bergstrom2015online, preibusch2016shopping, lee2013pibox, kokolakis2017privacy, bilge2013comparison}.

Compared with collaborative online shopping, both ridesharing and homesharing involve more interpersonal interactions (IPIs). For ridesharing, location privacy is separated from the general concept of privacy and is extensively studied and discussed. For homesharing, the general form of privacy is further divided into online privacy (electronic forms of personal information) and physical privacy (human body, house, furniture, etc.) \cite{christin2016privacy}. There are works showing that the physical privacy concerns are more important than online privacy concerns for homesharing \cite{belk2010sharing, lutz2017role}. We believe that the concept of physical privacy will be considered in privacy protection studies in other areas in near future, such as ridesharing.

It is noted that there are other methods available for privacy preservation in collaborative consumption practices in the early years. Milberg {\it et al.}~\cite{milberg1995values} studied various aspects that affected the customers' willingness to participate in collaborative consumption in the early 1990s. The study shows some early efforts and results from governments in designing suitable regulations for protecting the customers' privacy. Luo {\it et al.}~\cite{luo2002trust} examined several mechanisms to demonstrate the close relationship between trust and privacy preservation. Nissenbaum~\cite{nissenbaum2009privacy} discussed privacy from the perspective of contextual integrity in technology, policy, and social life.

\section{Summarizing Emerging Privacy Issues from the User, Platform and Service Provider Perspectives}\label{sec:emerging}

Fast development of cyber technology facilitates the invention of novel sharing practices in the cyber-enabled world. While traditional privacy problems have either been solved or at least realized by the government and society, privacy issues in cyber-enabled sharing services are less understood. In all six branches of the taxonomy in Figure \ref{fig:overall2}, there are always interactions between users, platforms, and service providers.

\begin{figure*}[h!t!b!]
  \centering
  \includegraphics[width=5.5in]{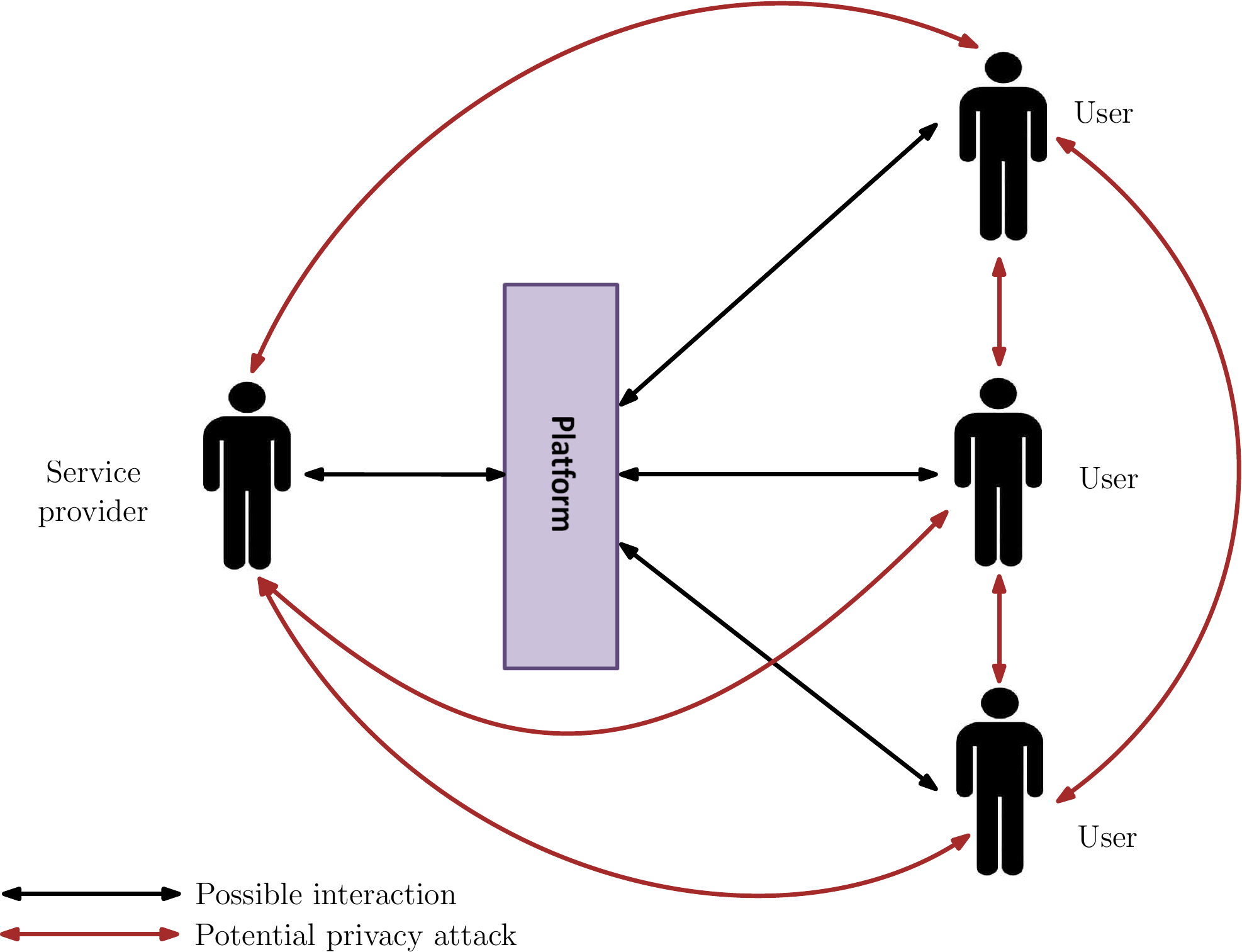}
    \caption{The privacy relationships between the user, platform and service provider. For any participant in the sharing
practice, no matter he/she is a user or service provider, all the remaining people involved in the same sharing practice can be potential attackers to compromise his/her privacy. }
    \label{fig:threePers}
\end{figure*}

\begin{figure*}[h!t!b!]
  \centering
  \includegraphics[width=6.5in]{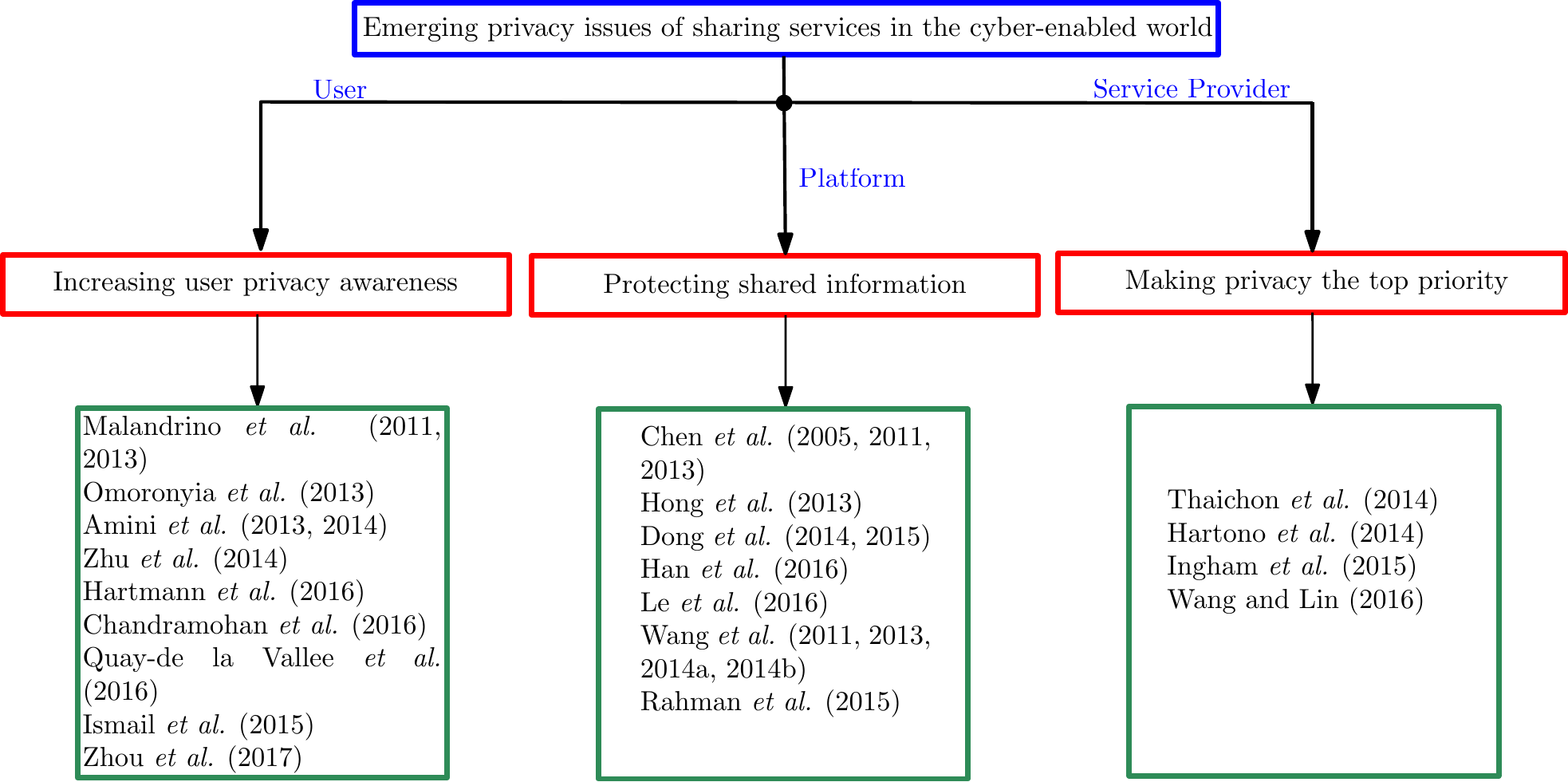}
    \caption{The emerging privacy issues identified in the current stage of cyberized sharing service development from the user, platform and service provider perspectives. The emerging privacy issues are shown in the red boxes. All works surveyed in this section are listed in the green boxes.}
    \label{fig:overall3}
\end{figure*}

In this section, all existing privacy issues are further discussed from three perspectives, namely, users', platforms' and service providers' perspectives. we argue that all privacy issues from different applications are internally related. Users concern with their own privacy and always demand high quality reliable sharing services. Service providers realize that privacy security level is a key element towards a successful achievement. Anybody involved in the sharing service can be a potential attacker to compromise other people's privacy. The linkages between the privacy concerns from the three perspectives are shown in Figure \ref{fig:threePers}. The concluded emerging issues of the cyber-enabled sharing services are: increasing users' privacy awareness from their perspective, protecting shared information from the platforms' perspective and making privacy concerns the top priority from the service providers' perspective. Works that are surveyed in this section are listed in Figure \ref{fig:overall3} and summarized from the three perspectives.

\subsection{From Users' Perspective: Increasing Privacy Awareness}\label{sec:emerging1}

Although most websites, software and mobile apps provide user agreements for user privacy awareness, only a negligible portion of users read through the tedious clauses carefully. The first emerging privacy issue for cyber-enabled sharing services is to maximize users' awareness of privacy leakage, e.g., to provide an online tool for users to trace down entities that may reveal their personal information. The transparent information tracing system will increase the confidence of users in participating in sharing practices on Internet, as well as facilitating the service providers to improve their reputations.

For example, in the crowdsourcing marketplace, it is not sufficient to protect only requesters' data privacy because workers also value their privacy equally. Workers are commonly afraid of the leakage of their location data or the identity information (e.g., age, contact, hobbies, activities)~\cite{wang2013respecting, to2014framework}. According to a survey that was performed by the U.S. Federal Trade Commission~\cite{federalprotecting},  more than $85\%$ of users were too impatient to read the user agreements regarding privacy settings carefully. They were surprised that mobile phone apps sent their approximate or precise location, phone's unique ID to service providers. Some apps even have control of the camera flashlight and audio settings. Although these privileges were authorized by users, they did not know when or where they give the authorizations, because they never read the articles about the privacy settings. Some efforts have been made to solve the above problem.

Malandrino {\it et al.} \cite{malandrino2011supportive, malandrino2013privacy} proposed a privacy awareness software named `NoTrace' to provide privacy recommendations to the users, such as privacy protection level settings, private information transmission warnings and unnoticeable privacy leaks warnings. The graphical user interface of `NoTrace' clearly displays the private information pieces that are received by the service provider. The main shortcoming of Malandrino {\it et al.}'s work is that they did not provide a deep analysis of which private information pieces are necessary for the service quality and therefore, could not provide proper recommendations on selective disclosure of personal information for users.

Omoronyia {\it et al.} \cite{omoronyia2013engineering} proposed an adaptive privacy framework to assist automatic privacy disclosure decision making for various applications. The framework is designed following the famous MAPE (Monitor, Analyse, Plan and Execute) loop, and is focused on three aspects: application attributes, potential privacy threats and derived benefits from privacy disclosure. One important insufficiency of their work is that it does not categorize privacy protection requirements according to different service functions, which makes the automatic privacy disclosure decision making relatively unreliable \cite{meis2016computer}.

Amini {\it et al.} \cite{amini2014analyzing,amini2013mobile} developed a software called `AppScanner' to help users better understand the functionalities of mobile applications. The software provides an informative description of what mobile apps are actually doing under a crowdsourcing environment. The transparency and detailed analysis of the mobile apps help make users aware of privacy leakage when using mobile apps for crowdsourcing. AppScanner only categorizes the mobile app behaviors as normal or abnormal. A detailed categorization according to the behaviors purposes, e.g., advertising and social networks, can be used to enhance the decision making ability for users \cite{wang2015using}.

Zhu {\it et al.} \cite{zhu2014mobile} implemented a mobile app recommendation system with security and privacy awareness. The proposed system first analyzes the mobile application with detection and diagnosis of the security risks from insecure data access permissions. The recommendation system then provides suggestions to the user on whether to continue using the mobile app according to the app's popularity and user settings. The recommendation is based on modern portfolio theory. The main insufficiency of Zhu {\it et al.}'s work is that the security risks are only evaluated based on the permissions that the mobile apps request.

Hartmann {\it et al.} \cite{hartmann2016privacy} summarized six main threats of mobile apps to make the users aware of potential privacy risks: insufficient control features, excessive data mining, data theft, surveillance, information leakage and social engineering. They also proposed eight recommendations for guarding against these privacy threats: privacy dashboard, privacy policy, data handling guidelines, user permissions, anonymization, IT infrastructure security, encryption,  and relationship. All the guidelines are valuable for future privacy-aware mobile application development. However, most importantly, immediate solutions for all conflicts are missing from both regulation and cyber technology perspectives.

Chandramohan {\it et al.} \cite{chandramohan2016new} concluded that over 90\% of users accept user agreements unconsciously, without knowing that their personal information can be misused. They described a complete privacy-preserving scheme called Petri-net Privacy-Preserving framework that was installed on a cloud server. However, the practicability and the scalability of their algorithm are still questionable.

Similar to traditional websites that force users to accept user agreements, the mobile apps mitigate the privacy risks to the users by requesting resource access permissions. Quay-de la Vallee {\it et al.} \cite{quay2016per} developed two app systems that help users find privacy-respective apps and manage the apps' permissions in their mobile phones. The main shortcoming of Quay-de la Vallee {\it et al.}'s work is that the two systems only provide privacy management assistance after the apps have been installed, instead of providing the assistance during the installations process.

Ismail {\it et al.} \cite{ismail2015crowdsourced} studied the privacy threats from mobile apps that require access to sensitive resources during the processes of installation or updating. A crowdsourcing strategy that identifies the minimal number of permissions to keep the mobile apps fully functioning for a diverse range of users was proposed. A user study that involved 26 participants and the popular mobile app `Instagram' showed the effectiveness of their approach. However, the survey size was relatively small; and the method was only tested on a single mobile app. The usability of the proposed crowdsourcing strategy requires further justification.

Zhou {\it et al.} \cite{zhou2017control} accessed the gap between users' desire of privacy control and the actual privacy setting functions provided by mobile app systems. Through a simple lab survey consisting of 26 users, three important facts had been concluded: 1) personal privacy protection is still an important factor that influences the users to choose their smartphones; 2) although smartphone nowadays provides more functions protecting user privacy through complex user interface, people are not well adapted to those new functions; and 3) Sorting methods, as well as recommendation systems are still useful to assist users to protect their private data. The shortcomings of Zhou {\it et al.}'s study is that the number of participated user is relatively small. Moreover, there's no specific solution has been proposed to increase the users' awareness of privacy protection.

\begin{table*}[h!t!b]\caption{References, main objectives, proposed solutions and important insufficiency of the surveyed works for increasing privacy awareness.}
\begin{center}
\begin{tabular}{p{2.8cm}p{0.8cm}p{3.6cm}p{3.6cm}p{3.6cm}}
 \hline \hline  \bf{Reference} & \bf{Year} & \bf{Main objective} & \bf{Proposed solution} & \bf{Important insufficiency}\\
 \hline {Malandrino {\it et al.} \cite{malandrino2011supportive, malandrino2013privacy}} &2011, 2013& {Measuring revealed data by service provider and privacy leakage to third-party websites} & `NoTrace' software & Lacking of analysis on necessary information disclosure for a known service\\
  {Omoronyia {\it et al.} \cite{omoronyia2013engineering}} &2013& {Assisting privacy disclosure decisions made by applications} & An adaptive privacy framework & Lacking systematical privacy requirements listing for a given set of service functions\\
  {Amini {\it et al.} \cite{amini2014analyzing,amini2013mobile}} &2013, 2014& {Helping users better understand the functionality of mobile applications} & AppScanner & A detailed categorization according to the behaviors purposes can be more helpful\\
  {Zhu {\it et al.} \cite{zhu2014mobile}} &2014& {Recommending mobile apps to users with security and privacy awareness} & A mobile app recommendation system &The security risks are only evaluated based on the permissions that the apps request\\
  {Hartmann {\it et al.} \cite{hartmann2016privacy}} &2016& {Addressing threats of mobile apps and proposing solutions} & Eight recommendations for six main threats & No immediate solution is provided\\
  {Chandramohan {\it et al.} \cite{chandramohan2016new}} &2016& Protecting user privacy on cloud & Petri-net Privacy-Preserving Framework & The practicability and real-time applicability of their algorithm need further discussion\\
  {Quay-de la Vallee {\it et al.} \cite{quay2016per}} &2016& Managing apps's access permissions & Two management apps & The privacy management assistance was only provided after the apps been installed\\
  {Ismail {\it et al.} \cite{ismail2015crowdsourced}} &2015& Identifying the minimal number of permissions to keep the mobile apps fully functioning & A crowdsourcing strategy & The proposed strategy is only tested on one single mobile app\\
  {Zhou {\it et al.} \cite{zhou2017control}} &2017& accessed the gap between users' desire of privacy control and the actual privacy setting functions provided by mobile app systems & A simple lab survey consisting of 26 users & No specific solution was proposed\\
 \hline \hline
\end{tabular}
\label{table:em1}
\end{center}
\end{table*}

In summary, all the above mentioned works, which we list in Table \ref{table:em1}, suggest that privacy leakage on some level is unavoidable for users to enjoy the sharing service. However, users' awareness of privacy leakage can be improved by listing threats from third-party websites/applications \cite{malandrino2011supportive, malandrino2013privacy, amini2014analyzing, amini2013mobile, hartmann2016privacy}, recommending safe decisions to users \cite{omoronyia2013engineering, zhu2014mobile} and using cyber-technologies \cite{chandramohan2016new, quay2016per, ismail2015crowdsourced}. Although various techniques are proposed to raise the users' awareness level, most sharing service platforms only provide user agreement terms to warn about possible privacy leakage. There is still a large gap between forcing users to agree to terms, granting access permissions to sensitive data and motivating users to actively protect their own privacy. Platform and service providers should be encouraged to use the existing cyber-technology to maximize users' awareness of privacy issues. Future works and surveys can be conducted in this direction.

\begin{figure*}[h!t!b!]
  \centering
  \includegraphics[width=6in]{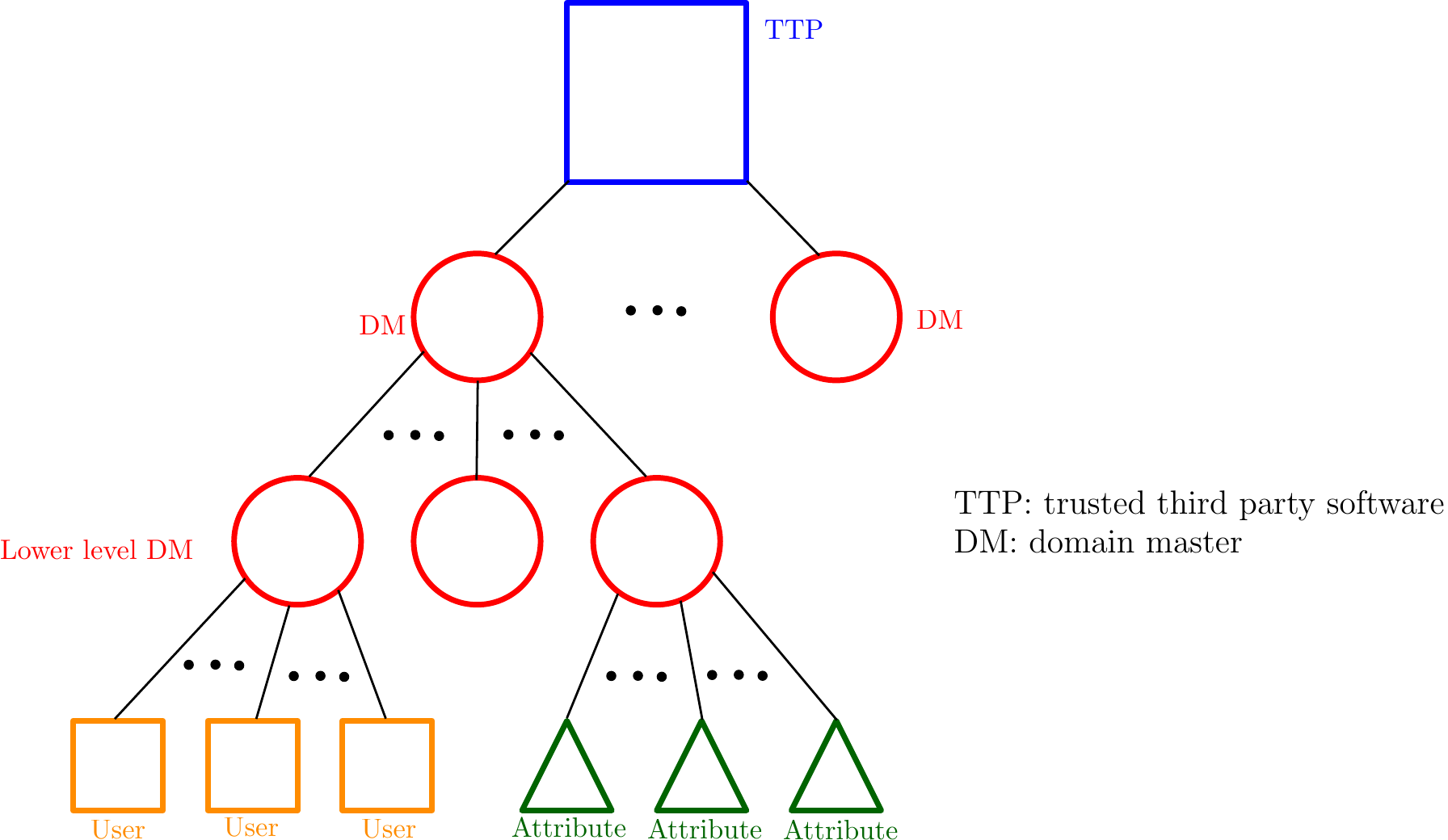}
    \caption{The hierarchical encryption scheme proposed by Wang {\it et al.}~\cite{wang2011hierarchical, wang2013secure, wang2014security, liu2014time}: the trusted third party has the access control for the domain masters. The domain master generates keys to a specific group of users in the next sub-level. For example, the leftmost domain master acts like the office administrate who is in charge of all personnel in the office, but not to administer any other attributes.}
    \label{fig:habe}
\end{figure*}

\begin{table*}[h!t!b]\caption{References, main objectives, proposed solutions and important insufficiency of the surveyed works for protecting shared user data.}
\begin{center}
\begin{tabular}{p{2.8cm}p{0.8cm}p{3.6cm}p{3.6cm}p{3.6cm}}
 \hline \hline  \bf{Reference} & \bf{Year} & \bf{Main objective} & \bf{Proposed solution} & \bf{Important insufficiency}\\
 \hline
  {Chen {\it et al.}~\cite{chen2005privacy, chen2011rasp, chen2013perturboost}} &2005, 2011, 2013& {Providing efficient and secure classifier using cloud computing technology with privacy preserved} & A random space encryption (RASP) approach & Updating the encrypted database is not an easy task\\
  {Hong {\it et al.}~\cite{hong2013survey}} &2013& {Preserving privacy under distributed environment} & Surveying existing privacy protection strategies & Mainly focusing on time-series data mining\\
  {Dong {\it et al.}~\cite{dong2014achieving, dong2015seco}} &2014, 2015& {Suggesting a privacy-preserving data security policy} & A series of encryption techniques & Resulting in key escrow problems\\
  Han {\it et al.}~\cite{han2016security} & 2016 & privacy-preserved data outsourcing under cloud environment & ABE based privacy protected data access control scheme & Requiring efficiency improvements\\
  {Le {\it et al.}~\cite{le2014consistency}} &2014& {Ensuring the enforceability for multi-access to stored data in cloud servers} & An inconsistency checking and removing algorithm & Requiring pre-defined rule regulations\\
  {Wang {\it et al.}~\cite{wang2011hierarchical, wang2013secure, wang2014security, liu2014time}} &2011, 2013, 2014& {Keeping the shared data confidential against untrusted cloud service providers} & The hierarchical attribute-based encryption scheme & lacking user revocation and was restricted by the same domain condition\\
  {Rahman {\it et al.}~\cite{rahman2015survey}} &2015& {Protecting shared data on cloud} & An information protection model combining incident handling strategy and digital forensics principles & The surveyed works were only up to the year 2014\\
 \hline \hline
\end{tabular}
\label{table:em2}
\end{center}
\end{table*}

\subsection{From the Platforms' Perspective: Protecting Shared Information}

Although users can agree to share part of their personal information on the intermediate platform, the shared information/data still faces various potential attacks without proper regulation protocol setups or cyber technology implementations. Data analysis for different purposes exists in almost all third-party platforms \cite{Reddy2018Comparative}. The main purpose of data analysis is to achieve better service quality. However, privacy concerns make users reluctant to share sensitive information. In this section, several recent existing works for privacy protection from the platforms' perspective are surveyed.

Chen {\it et al.}~\cite{chen2005privacy, chen2011rasp, chen2013perturboost} presented a random space encryption (RASP) scheme that produces secure privacy protection on the cloud. RASP provides service to transfer the analyzing data into an encrypted space with a two-stage encoding algorithm. The way of updating the encrypted database is another important challenge for their work.

Hong {\it et al.}~\cite{hong2013survey} surveyed several existing privacy protection strategies under the distributed data sharing environment. The proposed privacy protection techniques were simultaneously applied to the database, queries or aggregation. The main insufficiency of their work is that they only focused on privacy-preserving schemes for time series data processing.

Dong {\it et al.}~\cite{dong2014achieving, dong2015seco} suggested a security policy based on existing encryption techniques. The proposed framework allows the users to dynamically access their own personal data freely. Both attribute based encryption (ABE) and identity based encryption(IBE) were used to minimize the key management overhead; however, the proposed method resulted in key escrow problems \cite{sajid2016data}.

Following Dong {\it et al.}'s work, Han {\it et al.}~\cite{han2016security} provided a promising solution for privacy-preserved data outsourcing under the cloud environment. They proposed an attribute-based encryption (ABE) based control scheme on two major problems for data accessing privacy protection on the cloud. However, the time complexities of both the encryption and decryption processes in the proposed method were not optimized for real-world applications.

Le {\it et al.}~\cite{le2014consistency} assumed that there were pre-defined rule regulations in the data processing scenarios. An inconsistency checking and removing algorithm was designed to ensure the enforceability for multi-access to stored data in cloud servers. The main concern of their work is that the pre-defined regulations can be not applicable under extreme conditions or worst case scenarios.

Wang {\it et al.}~\cite{wang2011hierarchical, wang2013secure, wang2014security, liu2014time} proposed a hierarchical encryption scheme to maintain access controls for different levels of users (Figure \ref{fig:habe}). Each domain master generates keys to a specific group of users in the next sub-level. In addition, they also proposed a scalable revocation scheme for users to access their own personal data. The proposed scheme lacked user revocation and was only applicable to the situation that all attributes were administered by the same domain authority.

Rahman {\it et al.}~\cite{rahman2015survey} reviewed 139 works from 2009 to 2014 regarding information security in cloud computing. The cyber technology of incident handling strategy (IHS) is heavily discussed, which is an important tool for protecting data in a shared cloud service system. They pointed out that although IHS setup is straightforward on a personal computer, it becomes complicated when cloud computing allows multiple computers to access the same data on the same hard-disk. The main insufficiency of their work is that the survey was done in 2014 and only covered IHS techniques proposed before that year.

In summary, a list of the surveyed works can be found in Table \ref{table:em2}. From the platforms' point of view, there are mainly two parts of the data sharing practice can be worked on to provide more secure sharing services: the data transmission process and the data storage on the cloud server. To protect sensitive data during the data transmission process, data encryption is usually utilized \cite{dong2014achieving, dong2015seco}. For data protection on the cloud server, encryption scheme \cite{chen2005privacy, chen2011rasp, chen2013perturboost}, a hierarchical data-accessing scheme \cite{han2016security, le2014consistency, wang2011hierarchical}, and other cyber technologies \cite{rahman2015survey} were used. We believe that establishing an effective protocol in the platform is beneficial for both users and service providers. Although data analysis is necessary for service quality improvement, the part of the user data that must be revealed to the analyzer to obtain the full functionality of the sharing service remains questionable.

\begin{figure*}[h!t!b!]
  \centering
  \includegraphics[width=5in]{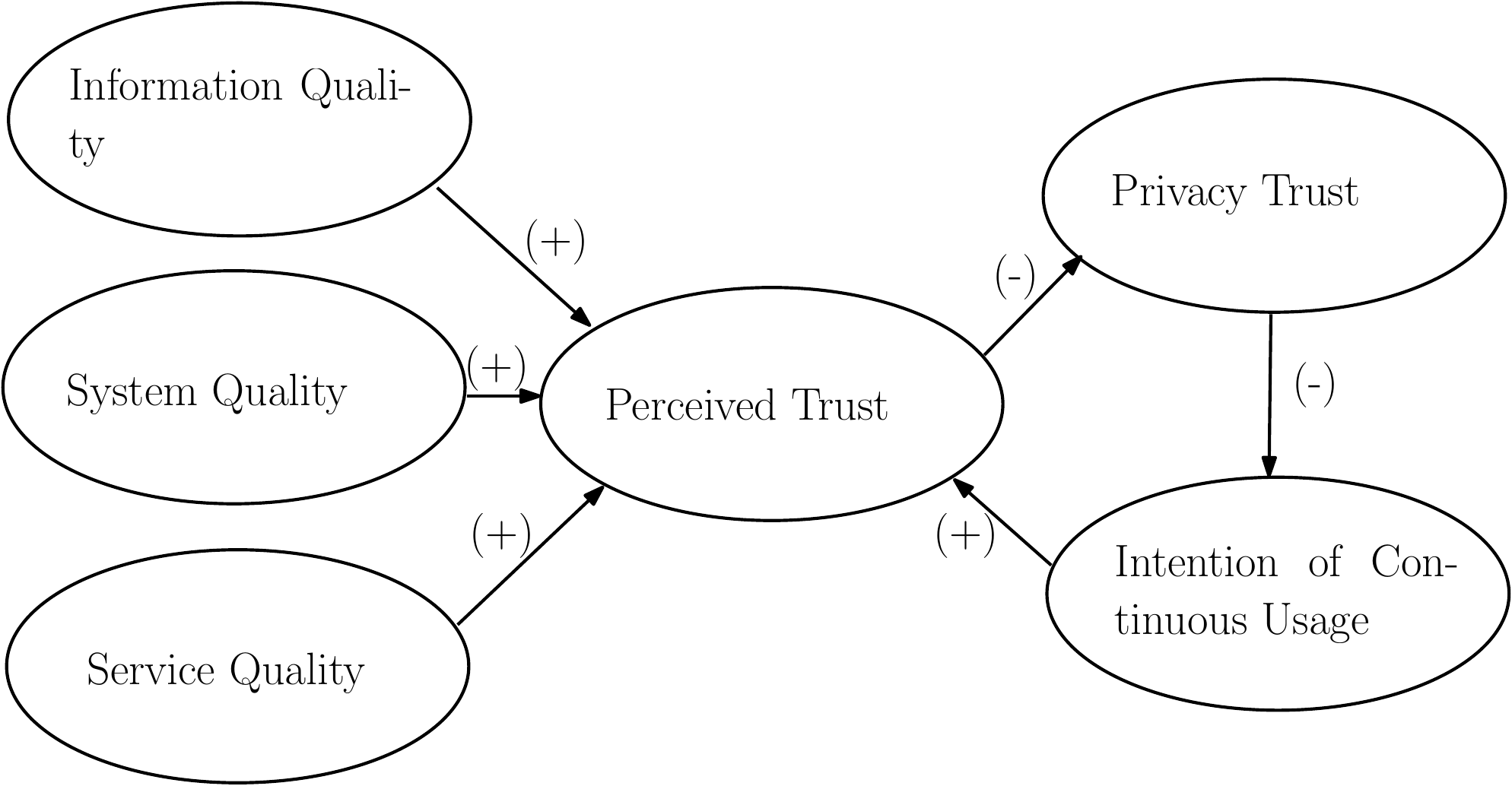}
    \caption{The research conceptual framework proposed by Wang and Lin's studies on the relationship between various elements on service quality and the intention of continuous usage of location based services~\cite{Wang2016Perceived}. The positive and negative influences between factors are marked by `+' and `-' signs.}
    \label{fig:wangLin}
\end{figure*}

\begin{table*}[h!t!b]\caption{References, main objectives, proposed solutions and important insufficiency of the surveyed works for realizing the importance of protecting user privacy.}
\begin{center}
\begin{tabular}{p{2.8cm}p{0.8cm}p{3.6cm}p{3.6cm}p{3.6cm}}
 \hline \hline  \bf{Reference} & \bf{Year} & \bf{Main objective} & \bf{Proposed solution} & \bf{Important insufficiency}\\
 \hline
  {Thaichon {\it et al.}~\cite{thaichon2014development}} & 2014 & {determining the relation between different service quality aspects (including privacy protection) and the final profit} & Identifying the four most important aspects for service quality enhancement & The survey results are only limited to a single country (i.e. Thailand)\\
  Hartono {\it et al.}~\cite{hartono2014measuring} & 2014 & Identifying the most important dimensions of perceived security for online shopping & A second-order structural model on perceived security & Only responses from Korea are used\\
  Ingham {\it et al.}~\cite{ingham2015shopping} & 2015 & Examining the internal relationship between trust, perceived risks,  and customers' acceptance & The technology acceptance model (TAM) nomological network & Lacking ways to gain the customers' trusts\\
  Wang and Lin~\cite{Wang2016Perceived} & 2016 & Studying the internal linking of service quality and intention of continuous usage of location-based services & A research conceptual framework & The survey was only conducted in Taiwan\\
 \hline \hline
\end{tabular}
\label{table:em3}
\end{center}
\end{table*}

\subsection{From the Service Providers' Perspective: Making Privacy the Top Priority}

As the last but important participant, the service provider has to learn the importance of protecting user privacy. Numerous studies have shown that privacy protection/security level is an important component of the overall service quality, and therefore influences the final profit of the company \cite{cases2010web, zhao2013designing, thaichon2014development}. More specifically, the enhancement of privacy protection quality by the service provider potentially attracts more customers to pay for the service \cite{carlos2009importance}. Service providers must give the privacy protection issue the highest priority in a successful business model.

Thaichon {\it et al.} \cite{thaichon2014development} surveyed the relationships between various aspects of service quality and the perceived value by customers. They identified the four most important service quality dimensions that influence the final profit of the company, which include privacy concerns. The limitation of their work is that the survey is conducted in the context of a single country (Thailand).

Hartono {\it et al.} \cite{hartono2014measuring} further identified the most important dimensions of perceived security for online purchases as confidentiality, integrity, availability,  and non-repudiation. They validated that these four aspects significantly impact the customer's willingness to participate e-commerce services by using a second-order structural model of perceived security. In their experiment, only responses from Korea were used, which reduces the generalization of the study results.

Ingham {\it et al.} \cite{ingham2015shopping} examined the internal relationships among trust, perceived risks and customers' acceptance in e-shopping practices. The technology acceptance model (TAM) nomological network is deeply discussed to measure the values in a different dimensions. The testing results are analyzed by the meta-analytical path approach. This was a comprehensive survey paper that searched for potential ways to promote e-commerce to achieve better sales. However, regulation or cyber technology solutions for enhancing the trusts gained from the customers are missing.

Wang and Lin \cite{Wang2016Perceived} established a conceptual research framework for studying the internal links between service quality and user experience of location-based services (LBS) (Figure \ref{fig:wangLin}). Based on a survey with 1399 participants, Wang and Lin identified positive and negative influences between factors, such as service quality and privacy trust in using LBS. Cultural bias exists in their results since the survey was conducted only in Taiwan.

All surveyed works from the service providers' perspective are listed in Table \ref{table:em3}. The internal relationship between the privacy protection and the net profit is heavily studied. The privacy protection level is an essential component in service quality evaluation and significantly impacts the customers' willingness to participate, customers' trust and net profit. And certain degrees of privacy disclosure from the service providers' side can also increase the willingness of the customers to trust the sharing services. In conclusion, it is important for the service providers to consider privacy issues the top priority of their commercial strategies, provide a more secure servicing environment and build more successful business models.

\section{Open Research Issues}\label{sec:openIssues}

In the first part of this study, the cyber-enable sharing services are divided into six branches, which are crowdsourcing marketplace, crowdfunding, crowdtesting, collaborative online shopping, ridesharing and homesharing. In this section, we summarize the main open research issues from the above six branches and list them as follows:
\begin{enumerate}
\item {\bf Improving task performance quality and efficiency with privacy-preserving protocols (crowdsourcing marketplace).} For Internet crowdsourcing marketplace, existing data manipulation approaches, such as coding theory and clipping protocols, decrease the task performance quality and efficiency. More efficient and effective mechanisms are demanded to better preserve the privacy from task requesters' perspective.
\item {\bf Degree of privacy sacrifice for the requesters towards a successful crowdfunding campaign (crowdfunding).} Trust is the key component for a successful crowdfunding campaign \cite{wen2018information, liang2018why}. However, the most appropriate degree of privacy sacrifice for the requesters remains as an open problem to attract more funding contributions.
\item {\bf Tradeoff between data encryption and testing result quality (crowdtesting).} Data encryption is a commonly used technique for protecting user privacy in crowdtesting practices, which unfortunately appear to decrease the testing result quality \cite{fabio2018understanding}. The way of balancing the tradeoff between data encryption and testing results quality is an important future working direction for crowdtesting practices.
\item {\bf An integrated approach to prevent misuse of customers' data (collaborative online shopping).} Data misuse is the main threat for customers who participate in the collaborative online shopping practice. Although there are solutions from both regulation and technical side, an integrated approach is demanded to better protect the users' privacy.
\item {\bf Conflict between location privacy and service based on location (ridesharing).} Location privacy is one of the hot topics in the field of location based services, such as ridesharing. However, there is always a conflict between hiding customers' real locations and utilizing the location information to serve customers better. A better solution to balance the conflict remains as an open problem in the field.
\item {\bf Physical privacy protection for hosts (homesharing).} For homesharing practices, existing works focus on mechanisms of protecting customers' privacy. However, from our study, homesharing involves lots of interpersonal interactions, where the physical privacy violation is also a potential threat for the hosts. A well-regulated scheme to better protect the physical privacy for hosts involved in homesharing practices remains open.
\end{enumerate}

In the second part of this work, the emerging privacy issues of the sharing services are further analyzed from three perspectives, namely, the users', platforms' and service providers' perspectives. The open problems from the three individual perspectives are:

\begin{itemize}
\item {\bf From users' perspective: motivating users to protect their own privacy.} While most of surveyed works use cyber technologies to protect users from potential privacy leakage, we pointed out that those techniques can only be used against unnoticeable threats. Utilizing cyber techniques to motivate the users to actively protect their own privacy is still the main solution and must be further emphasized in future works.
\item {\bf From platforms' perspective: establishing effective protocol for data analysis.} Encryption is a mature and commonly used cyber technique to protect user information during the data transmission and storage in sharing service platforms. Our study shows that, on top of the data encryption, a more sophisticated protocol is demanded for platform companies to access the necessary data for analysis in order for them to provide better services.
\item {\bf From service providers' perspective: enhancing the awareness of the importance of privacy protection using cyber technology.} From service providers' perspective, the surveyed works indicated that the privacy protection level is directly co-related to the net profit. However, the way of enhancing service providers' awareness for the importance of protecting users' privacy using cyber technology remains as an open problem for future studies.
\end{itemize}

\section{Conclusions}\label{sec:conclusion}

\begin{figure*}[h!t!b!]
  \centering
  \includegraphics[height=3.2in]{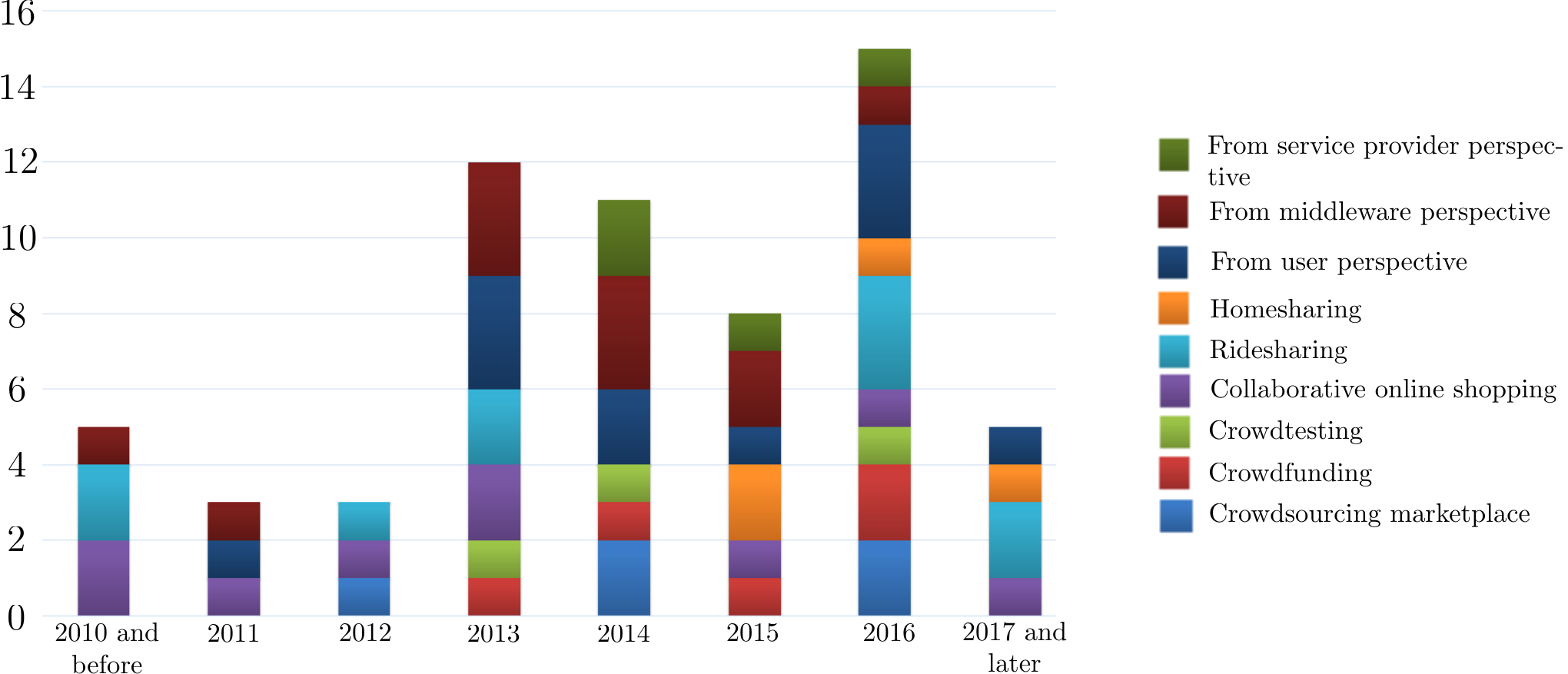}
    \caption{Yearly distribution of the number of all surveyed works from Table \ref{table:branch1} to Table \ref{table:em3}. Different colors are used indicating different types of sharing services.}
    \label{fig:statistics}
\end{figure*}

Privacy issues will sooner or later become the main barriers for both users and service providers who participate in the sharing economy. Over the past few years, great research efforts have been devoted to address various privacy issues existed in sharing service practices. Figure \ref{fig:statistics} shows the yearly distribution of the number of all surveyed works from Table \ref{table:branch1} to Table \ref{table:em3}. Different colors are used to indicate various types of sharing services. It can be clearly seen that a substantial part of the works published in the recent five years, i.e., starting from 2013 to 2017 and later, is surveyed in this study.

\begin{figure*}[h!t!b!]
  \centering
  \includegraphics[height=3.2in]{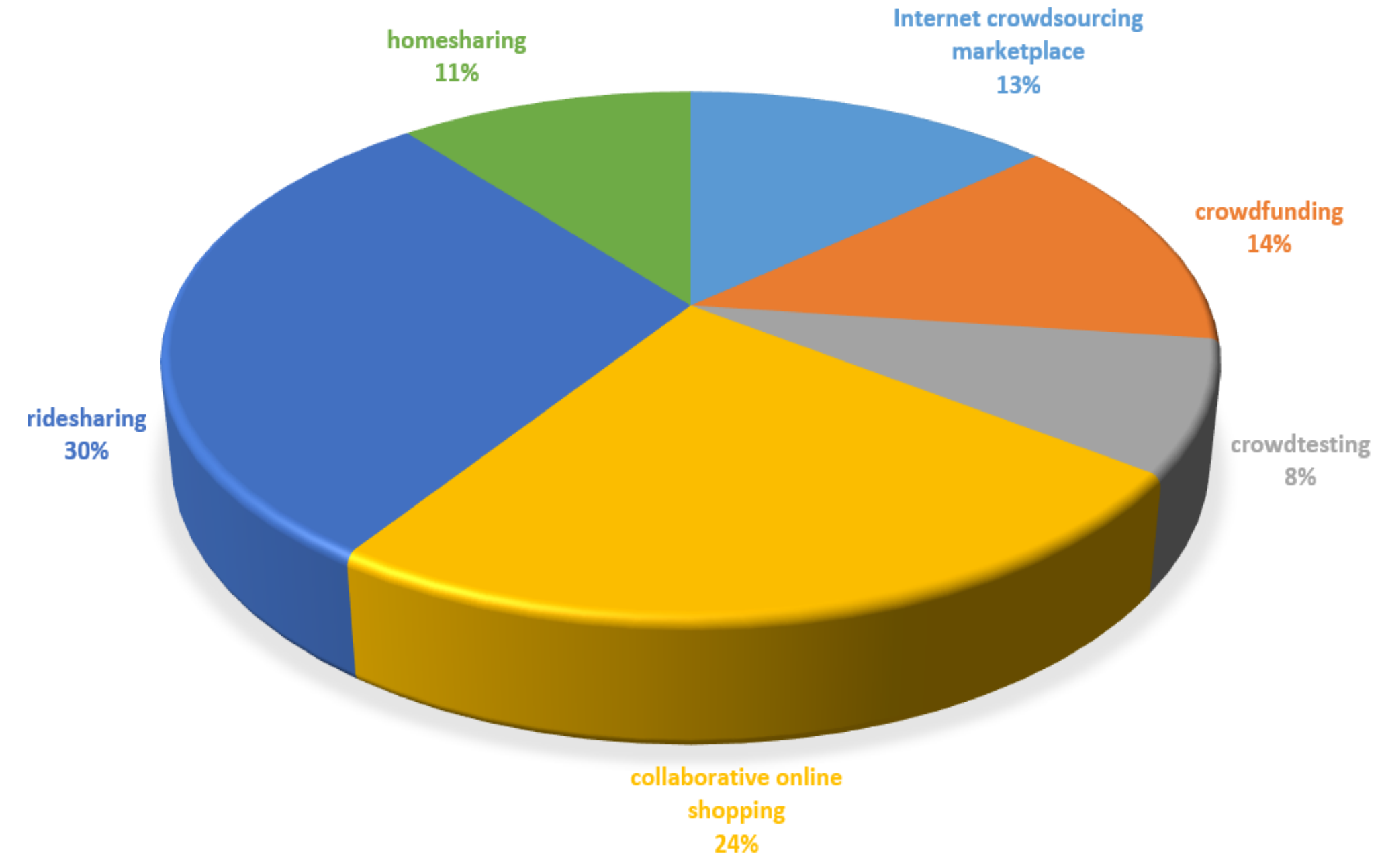}
    \caption{Statistical distribution of a total number of 37 works surveyed from Table \ref{table:branch1} to Table \ref{table:branch6}.}
    \label{fig:distribution}
\end{figure*}

\begin{figure*}[h!t!b!]
  \centering
  \includegraphics[height=2.8in]{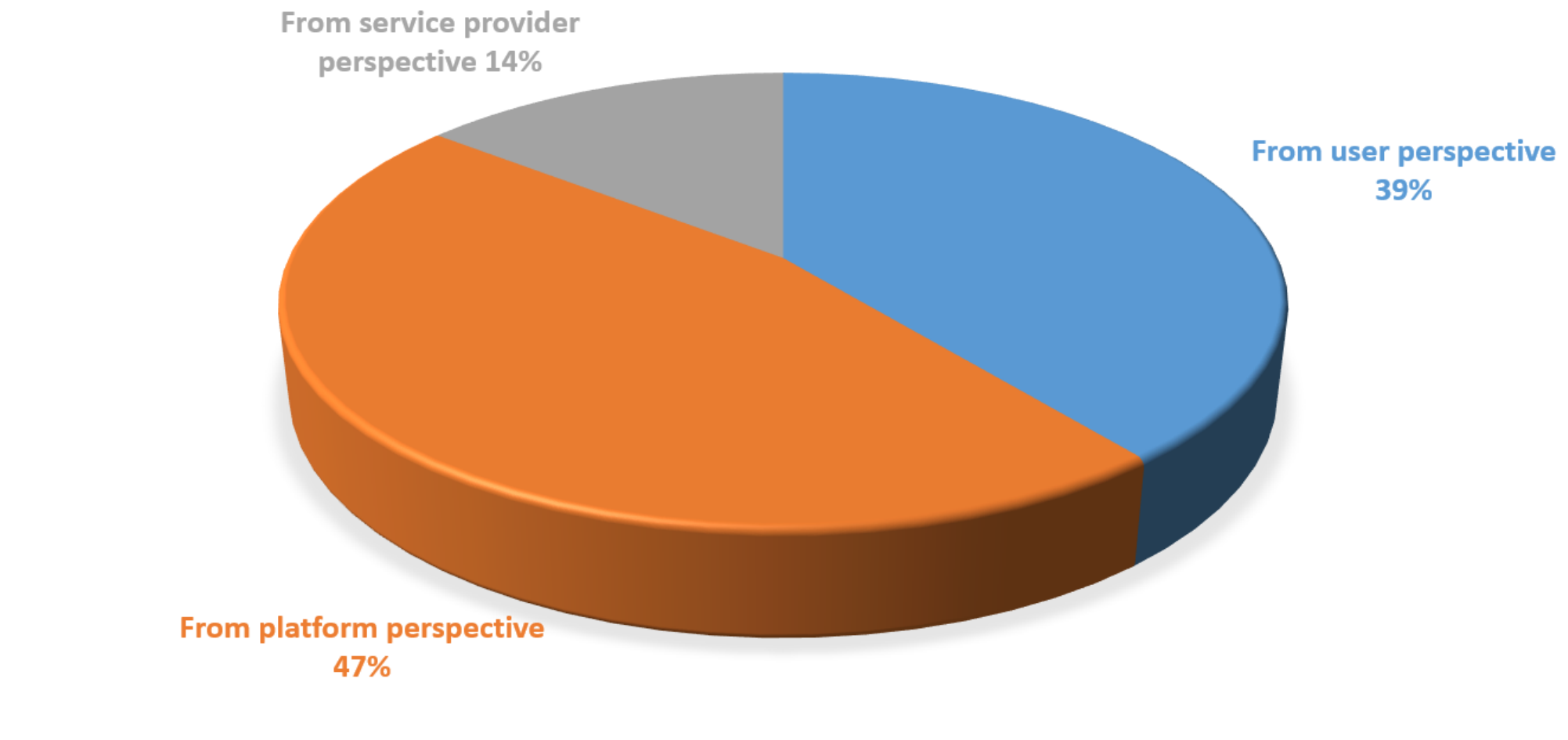}
    \caption{Statistical distribution of a total number of 28 works surveyed from Table \ref{table:em1} to Table \ref{table:em3}.}
    \label{fig:distribution2}
\end{figure*}

The cyber-enabled sharing services were divided into two categories: crowdsourcing and collaborative consumption. Crowdsourcing is further divided into three branches: Internet crowdsourcing marketplace, crowdfunding and crowdtesting. In Internet crowdsourcing marketplace practices, we tackled the privacy protection problem for task requesters. Two approaches were surveyed: the coding theory and the instance clipping protocol. In crowdfunding practices, modern crowdfunding platforms, such as Indiegogo, allow users to select their preferred security level and conceal their personal information privately, such as their names and contribution amounts. However, the surveyed works suggest that a certain level of privacy sacrifice can be helpful in crowdfunding practice. For crowdtesting practices, three real-world applications were surveyed, including shared data protection on a the cloud server \cite{harkous2014c3p}, online surveys \cite{kandappu2013exposing} and indoor site survey practice \cite{li2016privacy}. The main difficulties in protecting the privacy in crowdtesting practices are identified, which leads to one of the future research directions in the crowdtesting field. In collaborative consumption, the three sub-categories are: collaborative online shopping, ridesharing and homesharing. Collaborative online shopping, as a new generation of online shopping experience, raises two potential privacy concerns. The first privacy concern is the misuse of user data for marketing analysis, which can be prevented by refining government regulation \cite{preibusch2016shopping}, masking customers' data before sending them out \cite{bilge2013comparison} or separating communication channels on the cloud server \cite{lee2013pibox}. The second privacy concern is related to users' awareness of privacy leakage in online shopping, which was further discussed in later sections. In ridesharing practice, it is important to note that revealing the passenger's information, such as location, is necessary for the user to utilize the service. For homesharing, the surveyed works reveal that the hosts are actually more concerned about their privacy leakage than the travellers. Most of the privacy concerns are physical privacy issues.

In summary, Figure \ref{fig:distribution} shows the distribution of all listed surveyed works from Table \ref{table:branch1} to \ref{table:branch6}, including 37 works in total. In overall, the topics of privacy issues in collaborative online shopping and ridesharing are heavily discussed, whereas the topics of privacy issues in crowdtesting are less noticed. Although the surveyed works in this study do not include all works discussing the privacy issues of sharing services in the literature, the distribution reflects some aspects of the hotness/coldness of each mentioned topic, which provides potential directions to researchers for their future studies.

The above six branches of privacy concerns in the cyber-enabled sharing world are further summarized at the later part of this work from three perspectives. From the user perspective, users have started to realize that they have to sacrifice a certain degree of personal information to enjoy the sharing services. Therefore, the emerging issue is to increase the privacy awareness of the users. From the platform perspective, it is necessary for the third party platform to analyze the user's shared data to improve the service quality. The emerging issue from the platform perspective is to develop an effective protocol for identifying and protecting sensitive data during the transmission process, as well as the storage on the cloud server. From the service provider perspective, privacy must be recognized as the most important issue in the business model, which potentially impacts the perceived security and trust as well as the final profit.

Figure \ref{fig:distribution2} shows the distribution of all surveyed works from Table \ref{table:em1} to \ref{table:em3}, including 28 works in total. In overall, most existing works focus on privacy protection solutions from user and platform perspectives. There are only a few works mentioning that the privacy protection level can be improved by making the service providers realize the importance of protecting user privacy for their businesses. The privacy protection solution from the service providers' perspective deserves more attentions in future studies.

Table \ref{table:technicalCompare} covers the main cyber techniques surveyed in this work to protect privacy in sharing service practices. Each of these works was carefully evaluated to summarize its advantages/disadvantages compared with the remaining methods. All methods listed in Table \ref{table:technicalCompare} provide important solutions to protect privacy in different sharing service practices. Some of these methods can be more preferable under particular contexts or scenarios. For example, for privacy protection in crowdsourcing marketplace, SocialCrowd is preferred if the computational speed is not the main concern \cite{amor2016discovering}. Otherwise, a collusion network, proposed by Celis {\it et al.} \cite{celis2016assignment}, can be more preferable to minimize the privacy leakage.

\begin{table*}[h!t!b]\caption{Important cyber techniques surveyed in this work: a technical comparison.}
\begin{center}
\begin{tabular}{p{2.8cm}p{1.47cm}p{4.2cm}p{4.0cm}p{4.0cm}}
 \hline \hline  \bf{Method} & \bf{Reference} & \bf{Main technique} & \bf{Advantage} & \bf{Disadvantage}\\
 \hline
  {Coding theory} & {\cite{varshney2012privacy}, \cite{vempaty2014reliable}, \cite{vempaty2014coding}, \cite{wang2005distributed}} & {Add random perturbations to sensitive data} & {Hide sensitive information from the workers} & {Lose the task performance quality} \\
  {Instance clipping protocol (ICP)} & {\cite{little2011human}, \cite{chen2012shreddr}, \cite{kajino2014instance}} & {Clip the task into pieces} & {Allow each work only to access one clipped piece} & {Decrease the quality of the task results} \\
  {Collusion network} & {\cite{celis2016assignment}} & {Integrate ICP with three operations} & {ICP with minimal privacy leakage (better than ICP)} & {Information leakage from the workers' side} \\
  {SocialCrowd} & {\cite{amor2016discovering}} & {Clustering algorithms with heuristic function} & {Data leakage was effectively prevented} & {High computational complexity} \\
  {Encryption scheme for crowdtesting} & {\cite{kandappu2013exposing}, \cite{li2016privacy}, \cite{bilge2013comparison}} & {Add noises to the test results} & {Sensitive data cannot be revealed easily} & {Original data can be distorted} \\
  {$\pi$-box} & {\cite{lee2013pibox}} & {Develop a third-party app to control sensitive data transmission} & {Separated channels are designed to control data transmission} & {Not all paid apps support $\pi$-box} \\
  {Clustering $k$-anonymity (CK) scheme} & {\cite{yao2010clustering}, \cite{pan2013preserving}} & {Encrypt the user location information} & {Help ridesharing companies in protecting the user location privacy} & {Decrease location information resolution; lack protection on sensitive data transmission}\\
  {Anonymous mutual authentication (AMA) protocol} & {\cite{ni2016ama, ni2016privacy}} & {Develop AMA protocol for real-time navigation system} & {Conceal both customers and drivers' sensitive information (better than CK scheme)} & {A trusted third party middleware is required} \\
  {Two-stage auction algorithm} & {\cite{wang2018truthful}} & {Integrate $k$-anonymity scheme with $\epsilon$-differential scheme} & {Conceal users' locations in a more sophisticated way} & {Add Gaussian white noise to actual locations} \\

  {NoTrace} & {\cite{malandrino2011supportive, malandrino2013privacy}} & {Provide privacy recommendations to the users} & {Display the private information pieces received by service provider} & {Lack analysis on necessary information disclosure for a known service}\\

  {AppScanner} & {\cite{amini2014analyzing, amini2013mobile}} & {Let users better understand the functions of mobile apps} & {Provide informative description of what mobile apps do} & {Only categorize the mobile app behaviors as normal or abnormal}\\

  {Mobile app recommendation systems} & {\cite{zhu2014mobile}, \cite{quay2016per}} & {Analyze security risks and request resource access permissions from users} & {More complete system design with detailed permission levels (better than NoTrace and AppScanner)} & {Security risks are hard to be measured during the installation process}\\

  {Random space encryption (RASP) approach} & {\cite{chen2005privacy, chen2011rasp, chen2013perturboost}} & {Perform privacy protection on the cloud} & {Provide an encrypted space with a two-stage encoding algorithm on the cloud} & {Difficult to update an encrypted database} \\ \\
  {Attribute based encryption (ABE) scheme} & {\cite{dong2014achieving, dong2015seco}, \cite{han2016security}} & {Encrypt data on the cloud} & {Allow users to dynamically access personal data freely} & {Time complexities are not optimized for encryption and decryption processes} \\ \\
  {Hierarchical encryption scheme} & {\cite{wang2011hierarchical, wang2013secure, wang2014security, liu2014time}} & {Maintain access controls for different levels of users} & {Each domain master generates keys to next sub-level users (better than ABE scheme)} & {Not applicable to the situations that attributes were administered by different domain authorities} \\
 \hline \hline
\end{tabular}
\label{table:technicalCompare}
\end{center}
\end{table*}

In conclusion, we would like to point out that the solutions for emerging privacy issues in the cyber-enabled world include many different aspects, such as developing a more sophisticated encryption scheme for masking the user data, proposing a more reliable recommendation system for user privacy management, implementing a more secure transmission protocol and etc. All these issues/solutions represent the future research directions for privacy protection in sharing service practices.

\medskip
\noindent  {\bf Conflict of Interests}\\
All authors declare that there is no conflict of interest regarding the publication of this manuscript.\\

\section*{Acknowledgment}

This study is supported by National Natural Science Foundation of China (No. 61850410531), Zhejiang Provincial Natural Science Foundation of China under Grant No. LY19F020016 and National Natural Science Foundation of China (No. 61602431).

\bibliographystyle{ieeetr}
\bibliography{root2}

\end{document}